\documentclass[english,usenatbib]{mn2e}
\usepackage[]{fontenc}
\usepackage[latin1]{inputenc}
\usepackage{amsmath}
\usepackage{graphicx}
\usepackage{amssymb}
\usepackage[authoryear]{natbib}
\usepackage{ifpdf}
\usepackage{subfigure}

\newcommand{\msun}{\mbox{${\rm M}_\odot$}}

\newenvironment{acknowledgements}{\section*{Acknowledgements}}{}

\newcommand{\dd}{{\rm d}}

\newcommand{\id}{\ensuremath{\,\mathrm d}}
\newcommand{\diff}[3]{\ensuremath{\displaystyle\frac{\id^{#2}#1}{\id {#3}^{#2}}}}

\newlength{\timeswidth}
\newlength{\pluswidth}
\newcommand{\plustimes}{\ensuremath{%
\settowidth{\timeswidth}{$\times$}%
\settowidth{\pluswidth}{$+$}%
\addtolength{\timeswidth}{\pluswidth}%
+\hspace{-.5\timeswidth}\times%
}}

\title[Structure and evolution of high-mass stellar mergers]{Structure
  and evolution of high-mass stellar mergers}

\author[E. Glebbeek et al.]{
  Evert Glebbeek $^{1}$, 
  Evghenii Gaburov $^{2}$, 
  Simon Portegies Zwart,$^{3}$ and
  \newauthor
  ~Onno R.~Pols$^{1}$\\
$^1$ Department of Astrophysics/IMAPP, Radboud University Nijmegen, P.O.  Box 9010, 6500 GL Nijmegen, The Netherlands\\
$^2$ SURFsara, P.O. Box 94613, 1090 GP Amsterdam, The Netherlands \\
$^3$ Leiden Observatory, Leiden University, P.O. Box 9513, 2300 RA Leiden, The Netherlands\\
}

\begin{document}

\maketitle

\begin{abstract}
In young dense clusters repeated collisions between massive stars may lead
to the formation of a very massive star (above  $100\,\msun$). In the past
the study of the long-term evolution of merger remnants has mostly focussed
on collisions between low-mass stars (up to about $2\,\msun$) in the
context of blue-straggler formation. The evolution of collision products of
more massive stars has not been as thoroughly investigated. 
In this paper we study the long-term evolution of a number of stellar
mergers formed by the head-on collision of a primary star with a mass of
$5$--$40$ \msun{} with a lower mass star at three points in its evolution
in order to better understand their evolution.

We use smooth particle hydrodynamics (SPH) calculations to model the
collision between the stars. The outcome of this calculation is reduced to
one dimension and imported into a stellar evolution code. We follow the
subsequent evolution of the collision product through the main sequence at
least until the onset of helium burning.

We find that little hydrogen is mixed into the core of the collision
products, in agreement with previous studies of collisions between low-mass
stars. For collisions involving evolved stars we find that during
the merger the surface nitrogen abundance can be strongly enhanced.
The evolution of most of the collision products proceeds analogously
to that of normal stars with the same mass, but with a larger radius and
luminosity. However, the evolution of collision products that form with a
hydrogen depleted core is markedly different from that of normal stars with
the same mass. They undergo a long-lived period of hydrogen shell
burning close to the main-sequence band in the Hertzsprung-Russell diagram
and spend the initial part of core helium burning as compact blue
supergiants. 

\end{abstract}


\section{Introduction}
Dense stellar systems, such as the cores of star clusters or galactic
nuclei, are crowded places where stars frequently interact with each
other. In globular clusters, close two-body encounters may form
binaries \citep{1983Natur.301..587H}; furthermore, two-body encounters
can be close enough that two stars can come into physical contact
which can lead to a merger \citep{1976ApL....17...87H}. Galactic
nuclei, where stellar densities reach values in excess of millions of
stars per cubic parsec, also harbour stellar collisions. It therefore
appears that stellar mergers are natural events in dense
stellar systems and this has been demonstrated by several $N$-body
simulations \citep{1999A&A...348..117P, 2001MNRAS.323..630H,
2004Natur.428..724P}.

Stellar mergers provide a formation channel for non-canonical stars
that cannot be otherwise explained by the standard theory of star
formation and evolution, such as blue
stragglers that are observed in both open and globular clusters
\citep{1953AJ.....58...61S,1955ApJ...121..616J,1976ApL....17...87H,1995A&AS..109..375A,2004ApJ...604L.109P,2007A&A...463..789A}.
In a young star cluster stellar mergers
might be responsible for the formation of massive stars 
such as the Pistol Star in the Quintuplet Cluster
\citep{1998ApJ...506..384F}. Other massive stars, such as Sher
$25$ in the massive Galactic cluster NGC 3603, may have been
formed via binary mergers and mergers may contribute significantly to the
observed population of (rotating) massive stars
\citep{2012Sci...337..444S,2013ApJ...764..166D}. Along with the single
merger events, some star clusters, such as the Arches close to Galactic
centre or R136 in the Large Magellanic Cloud, are dense enough that runaway
stellar mergers can occur \citep{2004Natur.428..724P}.

Simulating stellar collisions has attracted considerable attention in the
past decade, mostly focussing on globular clusters with the aim of
explaining the formation of blue stragglers. 
Blue stragglers can be formed by stellar collisions or by mass
transfer in binary systems. Either of these mechanisms can dominate in one
particular cluster, or in different regions of the same cluster
\citep{2004MNRAS.349..129D}. $N$-body and Monte-Carlo simulations of
clusters show that both formation channels are necessary to reproduce the
observed blue straggler population
\citep{2001MNRAS.323..630H,2005MNRAS.363..293H,Chatterjee2013}.
Nevertheless, there are still many uncertainties regarding blue straggler
formation
\citep{2007ApJ...661..210L,2011MNRAS.415.3771L,2013MNRAS.428..897L,Sills2013}.

Dense stellar systems that are abundant in young massive stars, such
as the cores of young dense star clusters, are also a natural environment
for stellar mergers \citep{1999A&A...348..117P, 2008MNRAS.384..376G}. In
contrast to globular clusters, the colliding stars have masses that are
much larger than the average stellar mass of their environment. 
If the cluster is dense enough the same star may experience repeated
collisions in a so-called `merger runaway' \citep{1999A&A...348..117P}.
Such a merger runaway is triggered by the gravo-thermal collapse of the
cluster core \citep{1984MNRAS.208..493B} and can continue until the target
star leaves the main sequence \citep{1999A&A...348..117P}.
Only
recently have researchers begun to focus their attention on collisions
between more massive stars
\citep{2005MNRAS.358.1133F, 2006MNRAS.366.1424D, 2007ApJ...668..435S,
2008MNRAS.383L...5G, 2011ApJ...731..128A}. These
studies were focused either on global properties, such as mass loss, or
on the internal structure of collision products; yet very little is known
about the evolution of such objects. 
In earlier work \citep{2009A&A...497..255G} we have explored the impact of
mass loss on the evolution of the merger runaway. Here we focus on the
structure and evolution of single merger products.

In this paper we attempt to improve our understanding by
systematically carrying out collision simulations between massive
main-sequence stars and studying their further evolution. The aim of this
work is to understand the evolution of merger products formed by the
merger of two ordinary massive main-sequence stars as a function of stellar
ages and masses. This paper is structured
as follows. In Section \ref{sect:methods} we present the methods we use to
carry out this research, in Section \ref{sect:init_cond} we describe the
initial conditions for our simulations and in Section \ref{sect:results} we
present our results. A discussion of these results and our conclusions are
presented in Section \ref{sect:discussion}.

\section{Methods}\label{sect:methods}
\subsection{Hydrodynamic simulations}\label{sect:Hydro}
\begin{figure}
\begin{center}
\includegraphics[width=0.5\textwidth]{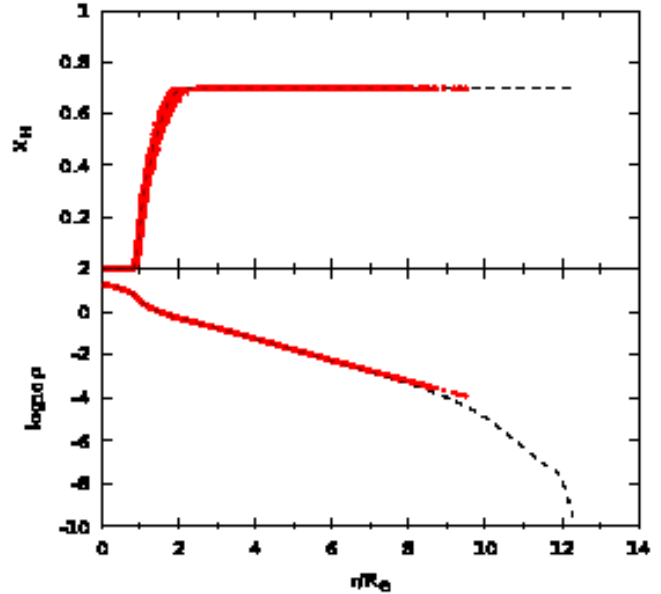}$\,$
\caption{Structure of the 20 \msun{} primary at the end of the main
sequence. The upper panel shows the hydrogen abundance and the lower panel
shows the density (in g/cm$^3$), both as a function of the enclosed radius.
Dots indicate individual SPH particles (155k, but only 1 in 10 is plotted), the
dashed line shows the one dimensional stellar evolution model.
}
\label{fig:CHEX_20Msun}
\end{center}
\end{figure}

We use smoothed particle hydrodynamics (SPH) to model stellar
collisions.
The full details of the SPH method are described in a number of review
papers, for instance \citet{Monaghan2005} or \citet{2009NewAR..53...78R}.
We repeat the main points here.
SPH is a fully Lagrangian method, which means that it easily adapts to any
geometrical configuration without some of the problems in finite-difference
methods, such as artefacts due to the choice of coordinate system or
numerical diffusion. In addition, the conserved fluid
quantities, such as composition, are trivially advected. The largest
drawback of SPH, however, is that low density regions are
poorly resolved. However, stellar interiors, which contain nearly all the
stellar mass, are dense enough for SPH to provide sufficient resolution to
capture fine details, such as the density and mean molecular weight
gradients across the boundary of a stellar core; for example, in Figure
\ref{fig:CHEX_20Msun} we show the structure of a 20 \msun{} star at the end
of the main-sequence phase.
Despite the fact that roughly 30\% of the stellar radius is unresolved,
this region contributes less than 1\% of the stellar mass, so that the
internal structure is well resolved.

The simulations that are presented in this paper are carried out by
means of a modified version of the {\tt GADGET2} code
\citep{2005MNRAS.364.1105S}.
In particular, we have modified the equation of state to include radiation
pressure.
The modifications are minimal and require only changes in the equation of state
and the equation for shock heating. The shock heating term becomes
\begin{equation}
  \frac{\id A}{\id t} = \frac{2}{3} \frac{\rho}{P\beta} A \frac{\id Q}{\id t}.
  \label{eq:dadt_dqdt}
\end{equation}
Here $dQ/dt$ is the SPH shock heating term which we have not modified,
$A(m)$ is the entropic variable (see section \ref{sect:pr_struct}) and $\beta$
is the ratio of gas pressure to total pressure.

\subsection{The stellar evolution code}\label{sect:Evolution}
We use the stellar evolution code originally developed by
\citet{eggleton_evlowmass} and later updated by others
\citep[\emph{e.g.}][]{pols_approxphys,GlebbeekPolsHurley2008}, hereafter {\tt STARS}.
The {\tt STARS} code solves the equations of stellar structure and the nuclear
energy generation rate simultaneously on an adaptive non-Lagrangian
non-Eulerian (``Eggletonian'') grid \citep{Stancliffe2006}.
Chemical mixing due to convection is treated as a diffusion process
\citep{bohm-vitense_convection,eggleton_mixingproc}, as is thermohaline
convection \citep{kippenhahn_thermohalinemixing, Stancliffe2007}.
Our mixing coefficient for thermohaline convection is based on the
expression of \citep{kippenhahn_thermohalinemixing},
\begin{equation}
D_\mathrm{thm} = C \frac{16 ac T^3}{c_P \rho^2 \kappa}
\frac{\nabla_\mu}{\nabla_\mathrm{ad} - \nabla}.
\end{equation}
The efficiency parameter $C$ is set to $100$ to reproduce the efficiency
calibration by \citet{2007A&A...467L..15C}.

Our reaction rates are the
recommended rates from \citet{NACRE}, with the exception of
$^{14}\mathrm{N} (\mathrm{p}, \gamma)^{15}\mathrm{O}$, for which
we use the recommended rate from \citet{Herwigetal2006} and
\citet{Formicola2004}.
Opacities are from \citet{IglesiasRogers1996}, with molecular opacities
from \citet{FergusonAlexander2005} and conductive opacities from \citet{Cassisi2007}.
Mass loss by stellar winds is included using the prescription of
\citet{vink_mass_loss, vink_mass_loss2} or \citet{dejager_mass_loss} at
cooler temperatures.

The conversion of the collision output to stellar evolution input is
done using the method described by \citet{GlebbeekPolsHurley2008}.
Once a suitable input model has been constructed we follow the subsequent
evolution of the merger remnant.

\section{Initial conditions}\label{sect:init_cond}
The aim of this work is to understand the structure and the evolution
of massive stellar collision products.
The structure of the collision product is influenced by the masses and
composition of the colliding (parent) stars as well as the parameters of
the collision.
We therefore need to span a section of the parameter space that encompasses
a variety of initial conditions for the geometry of the collisions and
the masses and ages of the parent stars.
In this study we focus on head-on collisions, ignoring the
complications associated with rotation in the merger product
\citep[\emph{e.g.}][]{2005MNRAS.358..716S} in order to focus on the
influence of the parent star masses and ages.
We consider only collisions where the total orbital energy is zero
(`parabolic' collisions).
This limitation can be
justified by the fact that the velocity dispersion in young star clusters is
much smaller than the escape velocity from the stellar surface and
therefore has little influence on the structure of the merger remnant.

\subsection{Masses and ages of the parent stars}
\begin{figure}
\includegraphics[width=0.5\textwidth]{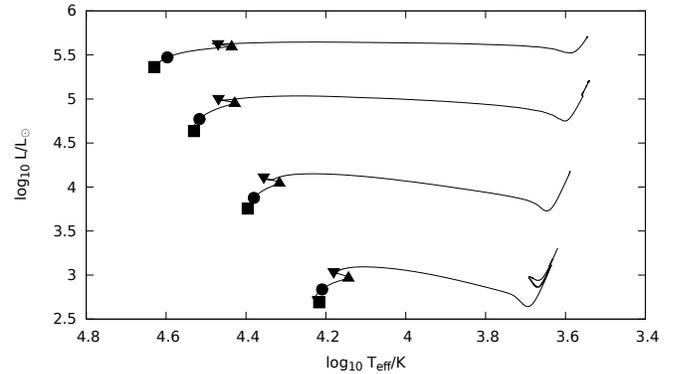}
\caption{
Hertzsprung-Russell diagram of our primary stars, with (from top to bottom)
masses of 40, 20, 10 and 5 \msun, showing the points chosen for our
collisions. The HAMS by $\bullet$
and TAMS and CHEX by $\blacktriangle$ and $\blacktriangledown$
respectively. For completeness, the ZAMS is indicated by $\blacksquare$.
}
\label{fig:hrd_primary}
\end{figure}

We systematically study collisions between massive stars of different
masses and ages, but with the same initial composition. We vary the mass of
the primary star $M_1$ and the mass ratio $q=M_2/M_1$ between the secondary
and the primary.

We choose the ages of the parent stars based on the evolution stage of the
primary, as illustrated in Figure \ref{fig:hrd_primary}: 
halfway through the main-sequence lifetime
(HAMS), at the  terminal-age main-sequence (TAMS) or at core hydrogen
exhaustion (CHEX). Here TAMS refers to the reddest point before the hook at
core hydrogen exhaustion. This normally corresponds to the moment where the star
has $3$--$4\%$ by mass of hydrogen left in the core, but in some of our
TAMS models the central hydrogen abundance is smaller ($1$--$0.1$\%). The CHEX
stage is shortly after this stage, at the bluest point after the TAMS stage
but before the Hertzsprung gap. This corresponds to actual core hydrogen
exhaustion, or a hydrogen abundance of about $0.01\%$ in the core depending
on the mass of the star.
The masses of the primaries are chosen to be $5$, $10$, $20$ and $40$
\msun, whereas the masses of the secondary star are chosen according to the
mass ratio $q$, which takes the values $0.1$, $0.4$, $0.7$ and $1.0$.
This was done to get a reasonable coverage between equal masses and extreme
mass ratios. For CHEX stars we choose $q = 0.99$ rather than $1.0$, which
roughly corresponds to a collision between a CHEX and a TAMS star.
To simplify the notation later we will refer to a collision between a 10
\msun{} TAMS star and a 1 \msun{} star (say) as `TAMS 10+1'.

\begin{table*}
  \caption{This table presents an overview of the simulations presented in this
    work. The first and second columns show the mass of the primary and
    the secondary star respectively, the third column shows the mass
    ratio. The fifth and sixth
    columns display the number of SPH particles used in each of the stars, and
    the last column indicates the evolutionary stage of the primary.
    In most of the simulations we use $262\mathrm{k}$ particles. One
    of the CHEX models is also simulated with higher resolution
    ($560\mathrm{k}$ particles) to verify that the lower resolution
    simulations capture the essential details. All stellar
    models are of solar metallicity ($Z = 0.02$).
    }
  \label{tab:simulations}
  \begin{tabular}{cccccc}
    \hline
    \hline
    $M_1$ [\msun] & $M_2$ [\msun] & $q$ & $N_1/1000$& $N_2/1000$ & Age \\
    \hline
    40 & 40   & 1.0  & 131  & 131  & TAMS, HAMS          \\
    40 & 39.6 & 0.99 & 132  & 130  & CHEX                \\
    40 & 28   & 0.7  & 155  & 107  & CHEX, TAMS, HAMS    \\
    40 & 16   & 0.4  & 400  & 160  & CHEX                \\
    40 & 16   & 0.4  & 187  & 75   & CHEX, TAMS          \\
    40 & 4    & 0.1  & 239  & 23   & CHEX, TAMS, HAMS    \\
    \hline
    20 & 20   & 1.0  & 131  & 131  & TAMS                \\
    20 & 19.8 & 0.99 & 132  & 130  & CHEX                \\
    20 & 14   & 0.7  & 155  & 107  & CHEX, TAMS, HAMS    \\
    20 & 8    & 0.4  & 187  & 75   & CHEX, TAMS          \\
    20 & 2    & 0.1  & 239  & 23   & CHEX, TAMS, HAMS    \\
    \hline
    10 & 10   & 1.0  & 131  & 131  & TAMS                \\
    10 & 9.9  & 0.99 & 132  & 130  & CHEX                \\
    10 & 7    & 0.7  & 155  & 107  & CHEX, TAMS, HAMS    \\
    10 & 4    & 0.4  & 187  & 75   & CHEX, TAMS          \\
    10 & 1    & 0.1  & 239  & 23   & CHEX, TAMS, HAMS    \\
    \hline
    5  & 5    & 1.0  & 131  & 131  & TAMS                \\
    5  & 4.95 & 0.99 & 132  & 130  & CHEX                \\
    5  & 3.5  & 0.7  & 155  & 107  & CHEX, TAMS, HAMS    \\
    5  & 2    & 0.4  & 187  & 75   & CHEX, TAMS          \\
    5  & 0.5  & 0.1  & 239  & 23   & CHEX, TAMS, HAMS    \\
    \hline
    \hline
  \end{tabular}
\end{table*}

\subsection{Stellar models and set-up of simulations}
The full set of simulations that we carried out is presented in
Table \ref{tab:simulations}. Most of the simulations use
$262\mathrm{k}$ equal mass SPH particles. This number was chosen
such that we are able to accurately resolve the internal structure of the
collision product \citep{2002MNRAS.332...49S,2008MNRAS.383L...5G}.

We prepare our 3D SPH models based on 1D stellar evolution models
calculated with {\tt STARS}.
Our input models are non-rotating and start with an initial heavy element
abundance (metallicity) $Z=0.02$ and a hydrogen abundance $X=0.70$.
For each collision we calculate the evolution of the primary from the
zero-age main-sequance (ZAMS) until the
appropriate age (HAMS, TAMS or CHEX) and then evolve the secondary
to the same age as the primary. We use the composition, density and
temperature profiles of the 1D models to construct a quasi-hydrostatic
equilibrium SPH model.

We use the resulting three-dimensional SPH models to prepare our collision
simulations. The parent stars are initially separated by a distance that is
equal to twice the sum of their radii and with the velocity of each star
computed in such a way that the total orbital energy, angular momentum and
velocity of the centre of mass is zero. The stellar velocities are
directed towards each other.

\subsection{Reduction of three-dimension data}
\label{sec:reduce_3d}
In order to import the results of our collision calculations into the
stellar evolution code the three-dimensional data has to be reduced to one
dimension and converted into a format that is understandable by the stellar
evolution code. 
A collision between stars is a complex hydrodynamic interaction of
self-gravitating fluids, and such interactions do not posses apparent
symmetries. Shocks together with turbulent heating result in mixing
of the fluid that is intrinsically three-dimensional. However,
the structure of a head-on collision product is spherically
symmetric once the fluid has settled into hydrostatic equilibrium. This
allows us to reduce the three-dimensional data to one dimension by
averaging over isobaric surfaces.

The collision calculations do not have sufficient resolution to resolve the
outer parts of the envelope near the photosphere, which means that we do
not have information about the structure of the envelope at that point.
However, for the stellar evolution code the input model needs to satisfy
the photospheric boundary condition. \citet{sills_evcolprod1} extrapolated
the entropy profile and used the condition of hydrostatic equilibrium to
reconstruct the outer layers.  Because our method to import the stellar
evolution models is fully implicit we find it easier to allow the evolution
code to simply adjust the unresolved layers in response to the stellar
interior \citep{GlebbeekPolsHurley2008}, which is equivalent to assuming that
the outer layers are in thermal equilibrium at the start of our evolution
calculations. This is a reasonable approximation because the thermal
timescale for these layers is short compared to the thermal timescale of
the entire star. As noted by \citet{sills_evcolprod1}, the long term
evolution of the collision product is not sensitive to the treatment of the
surface layers but the exact shape of the evolution track during the
contraction phase is.

\section{Results}\label{sect:results}
\subsection{Mass loss from the collision}
\begin{figure}
\includegraphics[width=0.5\textwidth]{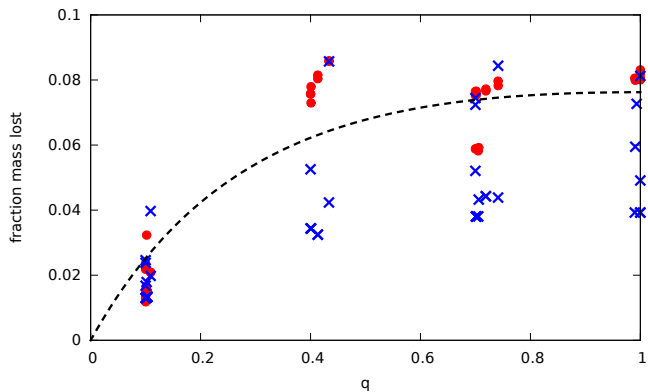}
\caption{
Mass loss from the collisions as a function of the mass ratio $q$. The
results from our simluations are indicated by $\bullet$, the
prediction of Eq. (\ref{eqn:massloss_lombardi}) is shown as $\times$
and the prediction of Eq. (\ref{eqn:massloss_simple}) is shown with
a dashed line.
}
\label{fig:massloss}
\end{figure}
For central collisions between low mass stars \citet{2002ApJ...568..939L}
found that the fraction of mass ejected by the collision $\phi$ can be modelled using
\begin{equation}\label{eqn:massloss_lombardi}
\phi = C_1 \frac{q}{(1+q)^2}\frac{R_{1,0.86} + R_{2,0.86}}{R_{1,0.5} + R_{2,0.5}}.
\end{equation}
Here $R_{n,0.86}$ and $R_{n,0.5}$ are the radii
containing 86\% and 50\% of the mass of parent star $n$ (1 or 2). The 
constant $C_1=0.157$. \citet{GlebbeekPols2008} found that for a set of
low mass collisions, the mass loss could also be modelled using the
simpler prescription
\begin{equation}\label{eqn:massloss_simple}
\phi = C_2 \frac{q}{(1+q)^2},
\end{equation}
with $C_2=0.3$. The result from our simulations is shown in Figure
\ref{fig:massloss} along with both of these prescriptions.
As can be seen from the figure, Eq. (\ref{eqn:massloss_lombardi}) gives a
better agreement for extreme mass ratio collisions but underpredicts the
mass loss at more equal mass ratios. The simpler prescription
Eq. (\ref{eqn:massloss_simple}) slightly overpredicts the mass loss at low mass
ratios but gives a better estimate at more equal masses.

\subsection{Structure of the collision products}\label{sect:pr_struct}

It was shown by \citet{2002ApJ...568..939L}
that there is a simple physical mechanism that determines
the structure of a collision product, and this provides a quick method
to obtain the approximate structure of the collision product as well as
to understand the outcome of SPH calculations. The
intricate details of shock and turbulent heating cannot be predicted by
simple analytical models, but an empirical tabulation of shock heating,
which is based on a number of simulations combined with conservation
laws, provides an accurate estimate of the degree of shock heating
\citep{2002ApJ...568..939L, 2003MNRAS.345..762L, 2008MNRAS.383L...5G}.

For a chemically homogeneous star in hydrostatic equilibrium the entropy
increases outward. The idea behind the method is to find a function of
entropy and composition that increases outward in stars that are not
chemically homogeneous.
Such a function is $A = P/\rho^{5/3} \exp [8(1-\beta)/3\beta]$
\citep{2008MNRAS.383L...5G}. 
We call this function $A(m)$ the buoyancy, since for stable hydrostatic
equilibrium the fluid with higher buoyancy should generally be above the
fluid with lower buoyancy. It is related to the entropy of the gas and is
also known in the literature as the entropic variable. The stability
condition is
\citep{2008MNRAS.383L...5G}
\begin{equation}
  \diff{\log A}{}{m} >
  \frac{4}{3}\frac{\frac{5}{3} - \Gamma_{1}} {\Gamma_{1} - \frac{4}{3}} 
  \diff{\log\mu}{}{m}.
  \label{eq:stability}
\end{equation}
Here $\mu$ is the mean molecular weight of the fluid element and
$\Gamma_1 = (\partial \log P / \partial\log\rho)_\mathrm{ad}$ is the
adiabatic index of the element
\citep{1983psen.book.....C,kippenhahn_weigert} and $m$ is the enclosed
mass.

In the case of a monatomic ideal gas ($\Gamma_1 = 5/3$) or a chemically
homogeneous fluid ($\dd\mu/\dd m = 0$) the
stability condition simplifies to $\dd A / \dd m > 0$. 
Heating is important because it modifies the entropy in each of the parent
stars. Therefore, for proper modelling one needs to take into
account the conversion of orbital kinetic energy into heat
\citep{2002ApJ...568..939L, 2008MNRAS.383L...5G}.

For collisions between unevolved (ZAMS) stars equation Eq. (\ref{eq:stability})
implies that the core of the lower mass star sinks to the centre of the
collision product. For evolved stars the situation is more complicated
because stellar evolution decreases $A$ in the core of the primary more
quickly than in the core of the secondary, but it is still possible that
the core of the secondary retains its identity and sinks to the centre of
the collision product. Following the results of \citet{GlebbeekPols2008}
for collisions between low-mass stars we identify the
cases `M', `P' and `S' depending on whether the core of the collision
product is a mixture of material from the progenitor stars, or
predominantly comes from the primary or the secondary.

\subsubsection*{Case M}
If the buoyancy in the cores of the two progenitor stars is similar,
or if the two stars are very close in mass, the material in the core
will be a mixture of the material in the cores of the two parent
stars. After the collision the hydrogen abundance in the core will be
in between the core hydrogen abundances in the progenitor stars. There can
be a molecular weight inversion if the material just outside the core
predominantly comes from the primary.

\subsubsection*{Case P}
If the primary is sufficiently evolved, the core of the primary becomes
the core of the collision product. This does not normally lead to a
molecular weight inversion, but it does mean that the collision product
has an anomalously small core for a star of its mass. If the core cannot
grow, for instance because hydrogen has already been exhausted and there is
no nuclear burning, the evolution path of the collision product can be very
different from that of a normal star of the same mass, as will be discussed
in section \ref{sec:evolution}.

\subsubsection*{Case S}
If stellar evolution has not decreased the buoyancy in the primary core
sufficiently, the core of the secondary will displace the core of the
primary and occupy the centre of the collision product. This can
happen at moderate mass ratios if the primary is relatively unevolved, or
it can happen at more extreme mass ratios even if the primary is already at
the end of the main sequence. In either of these cases the core of the
collision product will be hydrogen rich, with a helium rich layer just
outside the core.

\subsubsection{Half Age Main Sequence}
\label{sec:hams_structure}
Stars at half of their main-sequence age have converted a notable
amount of hydrogen into helium. During the collision, the stellar
interior is heated via shock waves, turbulent heating and tidal
interactions. The result is that some of the helium-enriched material
from the interior is mixed into the outer layers; at the same time, part of the
weakly bound outer layers are ejected. The amount of mass loss in these
collision is relatively small, as shown in
Table \ref{tab:mass_loss_hams}.

\begin{table}
\begin{center}
\caption{Mass lost from HAMS collisions. The first column is the ZAMS mass of
the primary star and its mass at the moment of collision in
brackets, the second column is the ZAMS mass of the secondary with the mass
at the moment of collision given in brackets, the third column is the
fractional mass loss as a percentage of the total mass and the last column
identifies the collision case (M, P or S, see text).
}
\label{tab:mass_loss_hams}
\begin{tabular}{cccc}
\hline
\hline
$M_1$ [\msun] & $M_2$ [\msun] & $M_{\rm lost}$ [\%] & Case  \\
\hline
40 (39.3)     &  40  (39.3)   &   6.2               & M \\
\hline
40 (39.3)     &  28  (27.7)   &   6.2               & M \\
20 (19.8)     &  14  (14.0)   &   6.5               & M \\
10 (9.98)     &  7   (7.00)   &   6.8               & M \\
5  (5.00)     &  3.5 (3.50)   &   6.9               & M \\
\hline
40 (39.3)     &  4   (4.00)   &   0.70              & S \\
20 (19.8)     &  2   (2.00)   &   0.66              & S \\
10 (9.98)     &  1   (1.00)   &   0.81              & S \\
5  (5.00)     &  0.5 (0.50)   &   1.3               & S \\
\hline
\hline
\end{tabular}
\end{center}
\end{table}

\begin{figure}
\begin{center}
\includegraphics[width=0.5\textwidth]{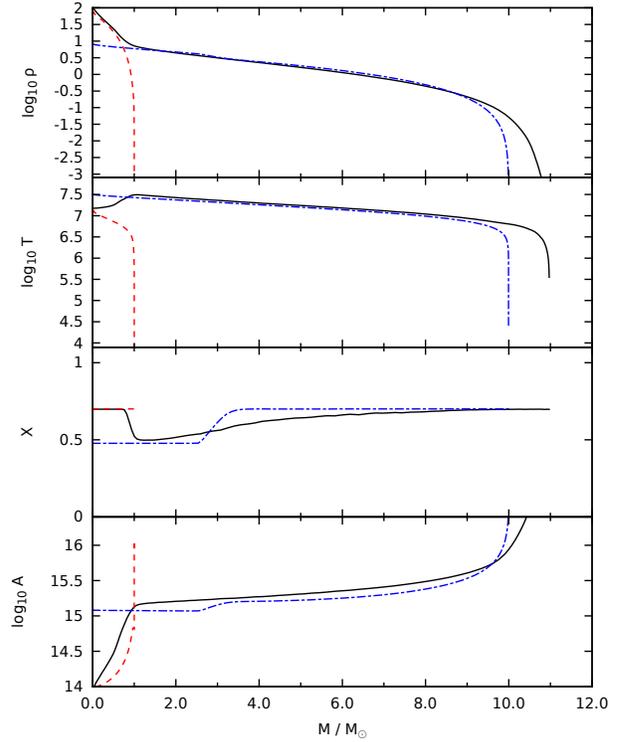}\\
\caption{Structure of the HAMS 10+1 merger product. The upper panel shows
the density profile as a function of
enclosed mass, the second panel from the top presents the
temperature profile and the two lowest panels display the hydrogen
mass fraction and buoyancy profile as a function of enclosed
mass respectively.
The dashed line (blue) shows the structure of the secondary and the
dash-dotted line (red) shows the structure of the primary.
}
\label{fig:hams10+1}
\end{center}
\end{figure}

In Figure \ref{fig:hams10+1} we show the structure of the HAMS 10+1
merger product, which is an example of case S that illustrates some of
the characteristic features.
The $1\,\msun$ star remains almost completely intact, as can be seen from
the hydrogen profile. This results in a molecular weight inversion at the
edge of the core at $1\,\msun$. Because the secondary retains its identity
but now finds itself compressed in the interior of a more massive star, the
core of the
merger product is overdense and overheated compared to a $1\,\msun$
star.
However, the entropy in the core is low compared to the entropy in a ZAMS
star of the same mass and composition as the merger product.
This results in a steepening of the density profile
below $1\,\msun$ and a temperature inversion, with the maximum temperature
occurring at 1 \msun.

\subsubsection{Terminal-age main sequence star (TAMS)}

\begin{table}
  \begin{center}
    \caption{As Table \ref{tab:mass_loss_hams} for TAMS collisions.}
    \begin{tabular}{cccc}
      \hline
      \hline
      $M_1$ (\msun) & $M_2$ (\msun) & $M_{\rm lost}$ (\%) & Case \\ 
      \hline
      40 (36.7)     &  40  (36.7)    &   8.3     & M \\
      20 (19.4)     &  20  (19.4)    &   8.0     & M \\
      10 (9.98)     &  10  (9.98)    &   8.1     & M \\
      5  (5.00)     &  5.0 (5.00)    &   8.2     & M \\
      \hline
      40 (36.7)     &  28  (27.2)    &   8.0     & M \\
      20 (19.4)     &  14  (13.9)    &   7.6     & P \\
      10 (9.98)     &  7   (7.00)    &   7.6     & P \\
      5  (5.00)     &  3.5 (3.5)     &   7.6     & P \\
      \hline
      40 (36.7)     &  16 (15.9)     &   8.6     & S \\
      20 (19.4)     &  8  (8.00)     &   8.2     & S \\
      10 (9.98)     &  4  (4.00)     &   7.3     & S \\
      5  (5.00)     &  2  (2.00)     &   7.5     & S \\
      \hline
      40 (36.7)     &  4   (4.00)    &   2.1     & S \\
      20 (19.4)     &  2   (2.00)    &   1.5     & S \\
      10 (9.98)     &  1   (1.00)    &   1.4     & S \\
      5  (5.00)     &  0.5 (0.50)    &   2.2     & S \\
      \hline
      \hline
    \end{tabular}
    \label{tab:mass_loss_tams}
  \end{center}
\end{table}

\begin{figure}
\begin{center}
\includegraphics[width=0.5\textwidth]{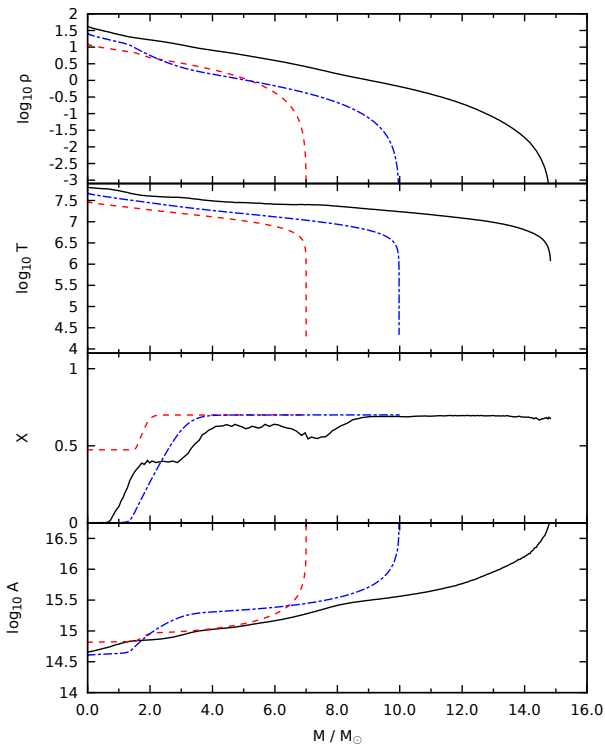}\\
\caption{As Figure \ref{fig:hams10+1} for the TAMS 10+7 merger product.
}
\label{fig:tams10+7}
\end{center}
\end{figure}

\begin{table}
  \begin{center}
    \caption{As Table \ref{tab:mass_loss_hams} for CHEX collisions.}
    \begin{tabular}{cccc}
      \hline
      \hline
      $M_1$ (\msun) & $M_2$ (\msun) & $M_{\rm lost}$ (\%)  & Case \\
      \hline
      40 (36.7)     &  39.6 (36.4)   &   8.0         & M \\
      20 (19.4)     &  19.8 (19.2)   &   8.0         & M \\
      10 (9.98)     &  9.9  (9.88)   &   8.0         & M \\
      5  (5.00)     &  4.95 (4.95)   &   8.1         & M \\
      \hline
      40 (36.7)     &  28  (27.2)    &   7.8         & M \\
      20 (19.4)     &  14  (13.9)    &   7.7         & P \\
      10 (9.98)     &  7   (7.00)    &   7.7         & P \\
      5  (5.00)     &  3.5 (3.5)     &   7.7         & P \\
      \hline
      40 (36.7)     &  16 (15.9)     &   8.6 (8.8)      & S \\
      20 (19.4)     &  8  (8.00)     &   8.2            & S \\
      10 (9.98)     &  4  (4.00)     &   7.8            & S \\
      5  (5.00)     &  2  (2.00)     &   7.5            & M \\
      \hline
      40 (36.7)     &  4   (4.00)    &   2.1         & S \\
      20 (19.4)     &  2   (2.00)    &   1.5         & S \\
      10 (9.98)     &  1   (1.00)    &   1.5         & S \\
      5  (5.00)     &  0.5 (0.50)    &   2.3         & S \\
      \hline
      \hline
    \end{tabular}
    \label{tab:mass_loss_chex}
  \end{center}
\end{table}

For TAMS stars the envelope is less strongly bound 
compared to HAMS stars and therefore a larger fraction of mass is lost, as
shown in Table \ref{tab:mass_loss_tams}. 
It can be seen that the mass loss fraction is also a decreasing function of
the mass ratio $q$. This is because the orbital kinetic energy, which is
dissipated into heat and used to eject the material, is proportional to the
product of the masses of the parent stars $M_1 M_2 = q M_1^2$.

An example of a case P merger product is the TAMS 10+7 product shown in
Figure \ref{fig:tams10+7}. In this case the core of the primary contained
$0.1\%$ by mass of hydrogen at the time of collision and the core of the
secondary is unable to displace the core of the primary. The inner
$0.5\,\msun$ of the remnant consists of hydrogen depleted material from the
primary, while the material between $0.5\,\msun$ and $1\,\msun$ is a
mixture of material from the primary and the secondary. The central density
and temperature are again higher than in the core of the $10\,\msun$
primary, but in this case there is no temperature inversion.

\subsubsection{Core Hydrogen Exhaustion (CHEX)}
\label{sec:chex_structure}
Stars that have completely exhausted hydrogen in their cores (CHEX stars)
have a similar structure to TAMS stars. Therefore we expect that the
structure of the merger products and mass loss are similar to TAMS
collisions. Indeed, Table \ref{tab:mass_loss_chex} shows that the mass lost
in collisions between CHEX stars is quantitatively similar to the mass lost
in collisions between TAMS stars.

\begin{figure*}
  \begin{center}
\subfigure[20 + 19.8]{\label{fig:chex20_198}\includegraphics[width=0.5\textwidth]{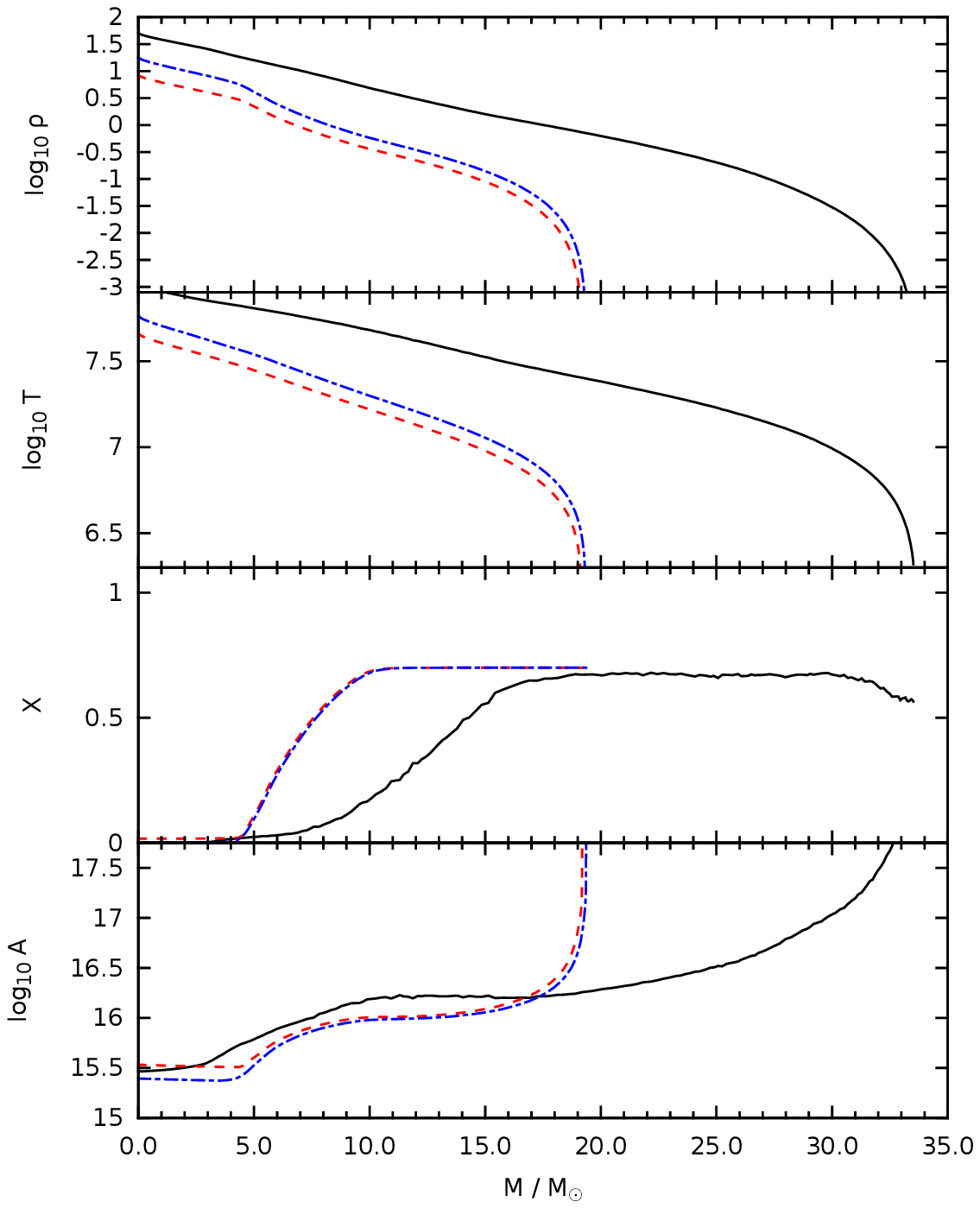}}%
\subfigure[20 + 14]{\label{fig:chex20_14}\includegraphics[width=0.5\textwidth]{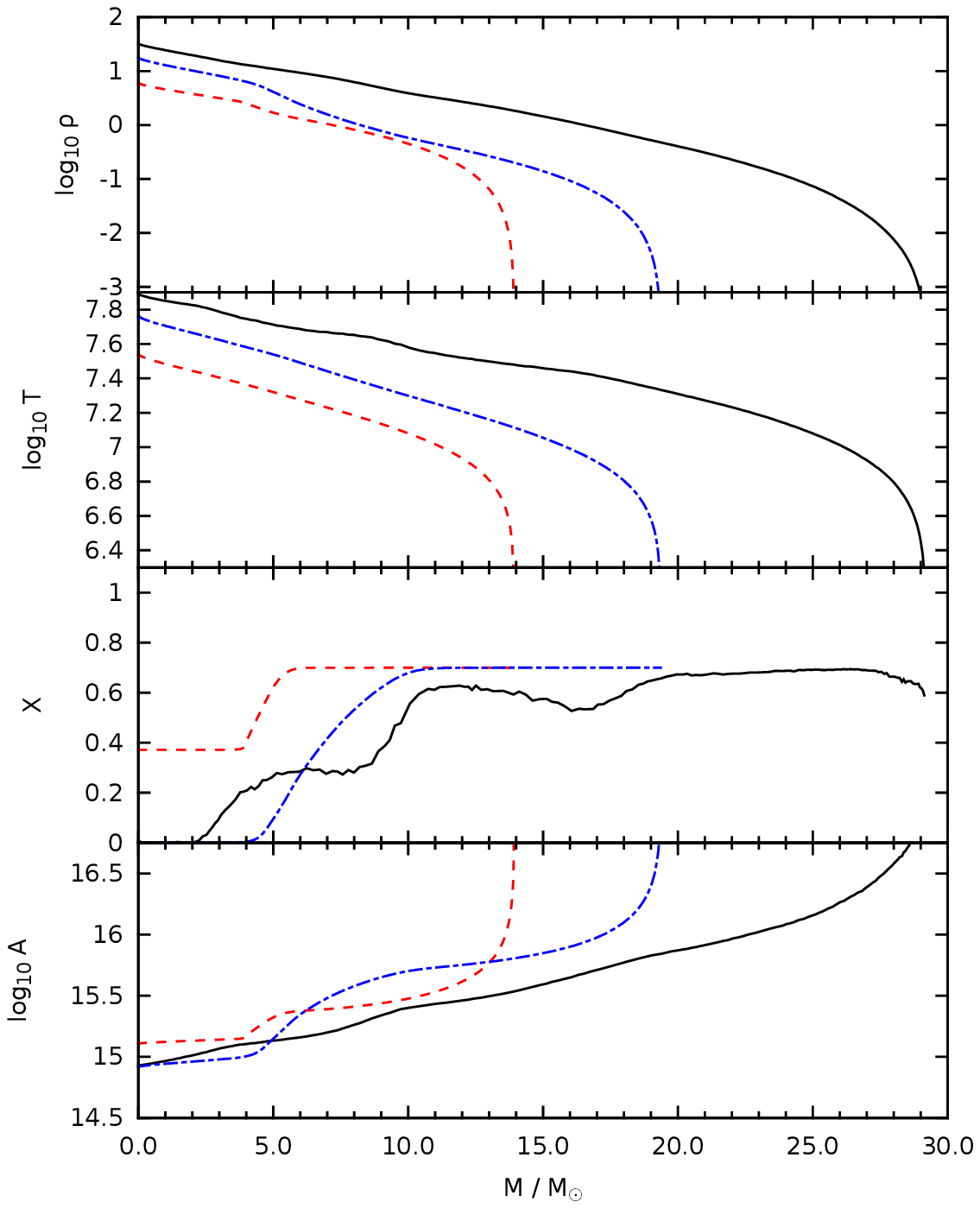}}

\subfigure[20 + 8]{\label{fig:chex20_8}\includegraphics[width=0.5\textwidth]{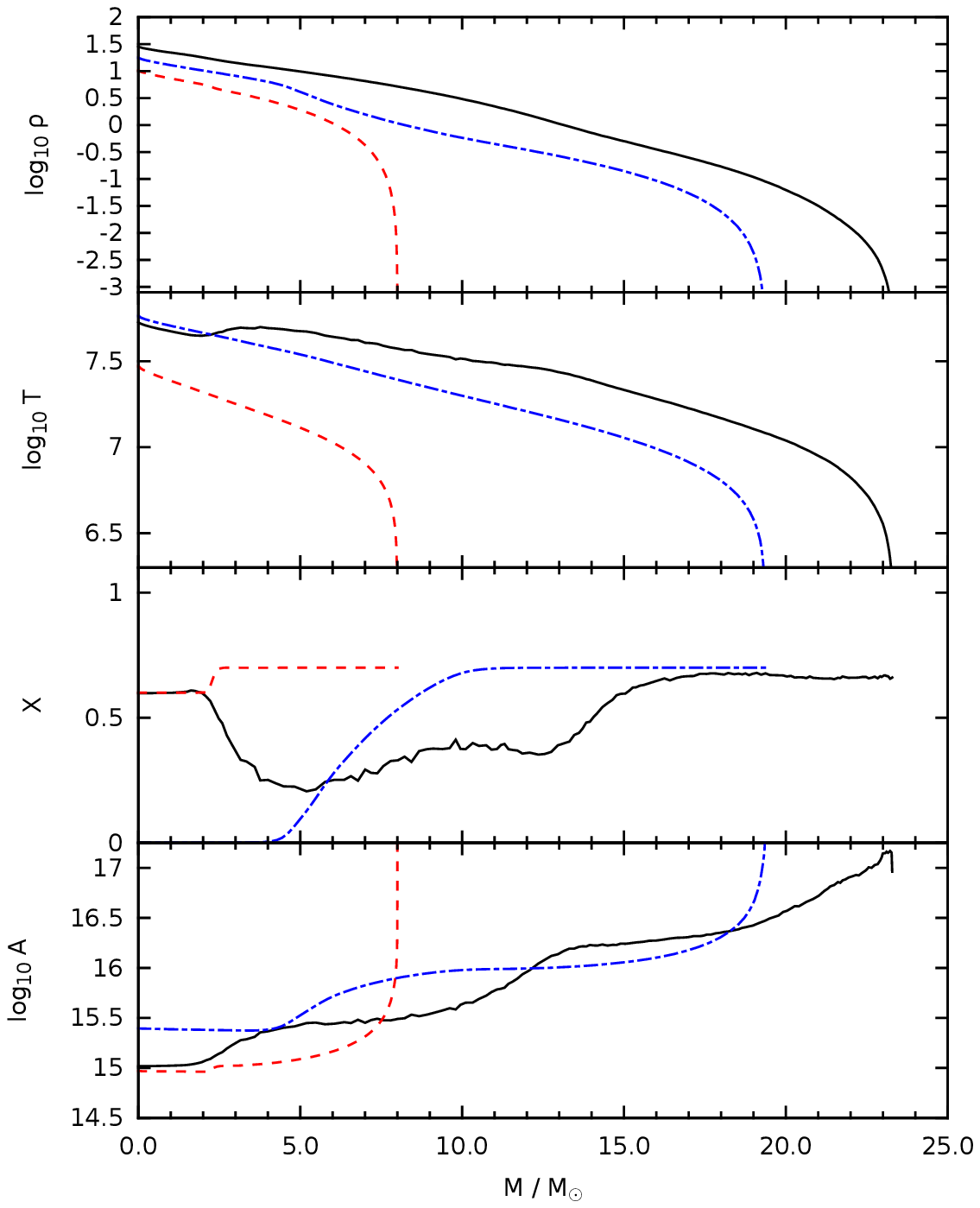}}%
\subfigure[20 + 2]{\label{fig:chex20_2}\includegraphics[width=0.5\textwidth]{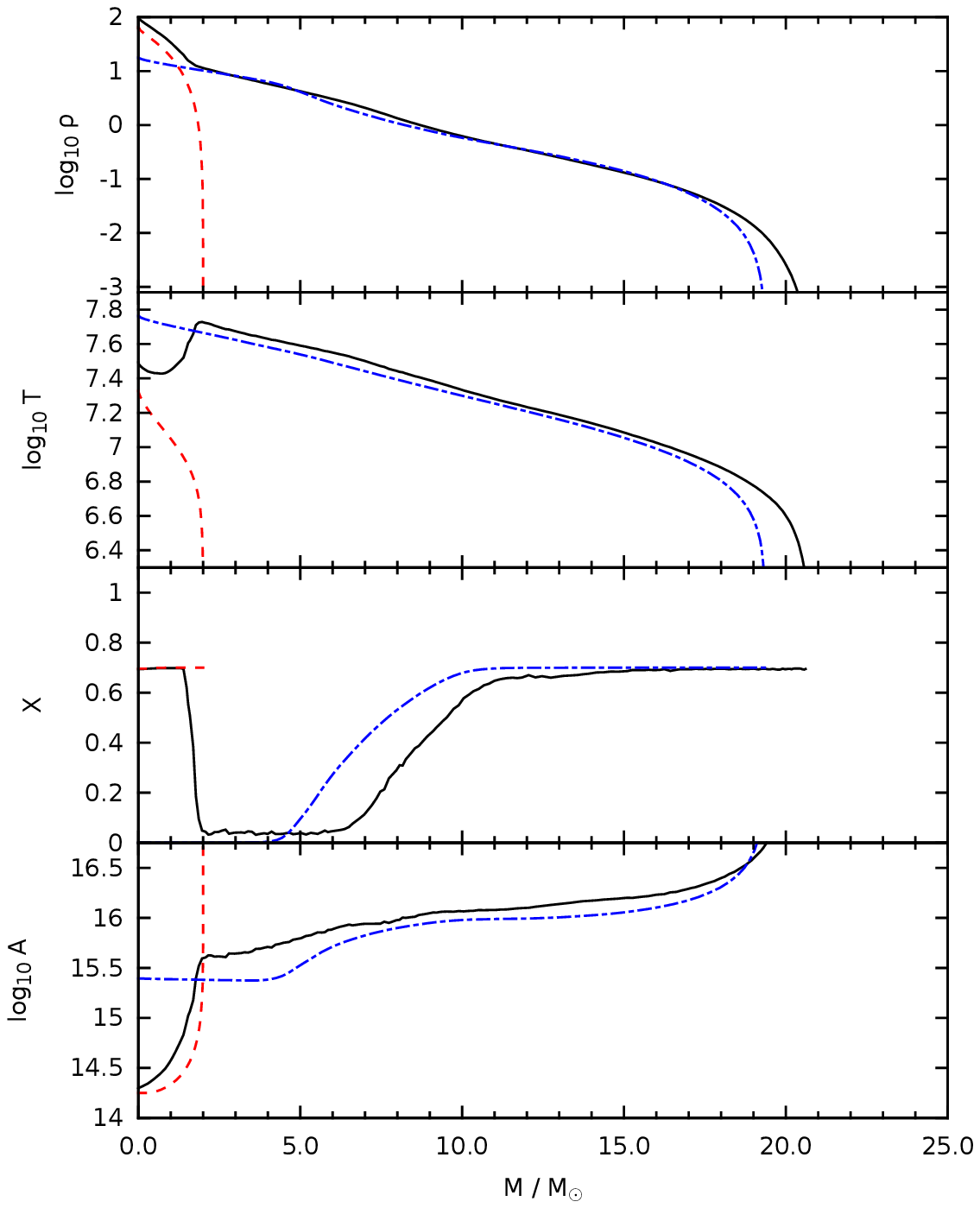}}

\caption{As Figure \ref{fig:hams10+1} for the CHEX collisions with a 20
\msun{} primary and 19.8 (upper left panel), 14 (upper right panel), 8
(lower left panel) and 2 (lower right panel) \msun{} secondaries.
In all cases the dashed line (red) shows the structure of the secondary
and the dash-dotted line (blue) shows the structure of the primary.
}
\label{fig:chex20}
\end{center}
\end{figure*}

In Figure {\ref{fig:chex20}} we show the structure of the merger products
of which the primary is a 20 \msun{} CHEX star and the secondary covers a mass
range such that the mass ratio $q$ is 0.99 (\ref{fig:chex20_198}), 0.7
(\ref{fig:chex20_14}), 0.4 (\ref{fig:chex20_8}) and 0.1
(\ref{fig:chex20_2}). 
Each of the three cases occurs in one of these panels.
The CHEX 20+19.8 collision is a clear example of case P, while CHEX 20+14
can be considered case M since a notable amount of hydrogen from the
secondary has been mixed into the core. Despite this, the material in the
core predominantly comes from the primary and is hydrogen poor.
The CHEX 20+8 and CHEX 20+2 are both case S, although the differences are
quite interesting.
In both cases the inner $\sim 2\,\msun$ consists of hydrogen-rich material
from the core of the secondary. In the CHEX 20+2 case, this means the
entire star occupies the core of the merger product with a strong
enhancement in the helium abundance of the envelope. In the CHEX 20+8 case
there is an additional $6\,\msun$ of hydrogen-rich material that is mixed
into the envelope, which dilutes the helium enhancement considerably.
As with the HAMS 10+1 merger product, the CHEX 20+2 remnant shows a
notable temperature inversion just outside the core and a clear kink in the
density profile. No such kink appears in the CHEX 20+8 case, although there
is a small temperature inversion. We will return to these features when
considering the evolution in section \ref{sec:contraction_and_mainsequence_evolution}.

\subsection{Evolution of the merger remnants}\label{sec:evolution}
We have calculated the evolution of the merger products 
listed in Table \ref{tab:simulations}.
The results of our evolution
calculations are presented in Table \ref{tab:evolution_results}.
A few models are missing from this table: we were unable to follow the
evolution of these models for more than a few timesteps before the
evolution code broke down. Also missing is the Hertzsprung gap evolution
for CHEX 5+0.5, which broke down shortly after the contraction phase.

\begin{table*}
\begin{minipage}{\textwidth}
\caption{%
Evolution results for the merger products. The first three columns list
the evolutionary stage of the primary (HAMS, ZAMS or CHEX), the mass of the
primary $M_1$ and the mass of the secondary $M_2$ (both in \msun). Column
four identifies the collision case (M, S or P, see text) and column five gives
the total mass of the remnant $M_\mathrm{rmn}$ at the beginning of the
evolution. Columns six through thirteen give the amount of mass loss in the
collision $\Delta M$, the total main sequence lifetimes $\tau_\mathrm{ms, 1}$
and $\tau_\mathrm{ms, 2}$ of the progenitor stars, the main sequence
lifetime of a star born with the same mass as the merger product
$\tau_\mathrm{ms}$, the actual main sequence (core hydrogen burning) lifetime $t_\mathrm{ms}$ of
the merger product, the time $\tau_\mathrm{HG}$ spend in the Hertzsprung
gap by a star born with the same mass, the time $t_\mathrm{HG}$ the
merger product spends in this part of the Hertzsprung-Russell diagram
and the time $t_\mathrm{HSB}$ between the end of core hydrogen burning
and central helium ignition (i.e.\ the length of hydrogen shell burning).
All times are given in Myr and masses are in solar units. The final two columns give the helium core mass
$M_\mathrm{He}$ at the moment of helium ignition and helium core mass
$M_\mathrm{He,ms}$ in a normal star at helium ignition.
} \label{tab:evolution_results}
\begin{tabular}{rrrrrrrrrrrrrrr}

\hline
Stage & $M_1$ & $M_2$ & Case & $M_\mathrm{rmn}$ & $\Delta M$ &
$\tau_\mathrm{ms, 1}$ & $\tau_\mathrm{ms, 2}$ & $\tau_\mathrm{ms}$ & $t_\mathrm{ms}$ & $\tau_\mathrm{HG}$ & $t_\mathrm{HG}$ & $t_\mathrm{HSB}$ & $M_\mathrm{He}$ & $M_\mathrm{He, ms}$\\
\hline
HAMS &    5 &  0.5 & S &  5.43 &   0.07 &   82.03 &  124027 &   81.92 &   57.43 &    0.52 &    0.52 &    0.64 &  0.98 &  0.94\\
HAMS &   10 &    1 & S & 10.86 &   0.13 &   19.82 &    8646 &   19.87 &   13.44 &    0.07 &    0.07 &    0.07 &  2.67 &  2.53\\
HAMS &   20 &    2 & S & 21.43 &   0.36 &    7.72 &     908 &    7.98 &    5.25 &    0.03 &    0.03 &    0.02 &  6.55 &  6.19\\
HAMS &   40 &    4 & S & 41.90 &   1.40 &    4.28 &     141 &    4.48 &    2.81 &    0.01 &    0.01 &    0.01 & 16.78 & 15.34\\
HAMS &    5 &  3.5 & M &  8.00 &   0.50 &   82.03 &     199 &   35.26 &   27.38 &    0.19 &    0.13 &    0.15 &  1.70 &  1.60\\
HAMS &   10 &    7 & M & 16.00 &   0.99 &   19.82 &   38.89 &   11.17 &    8.43 &    0.06 &    0.04 &    0.02 &  4.24 &  3.91\\
HAMS &   20 &   14 & M & 31.76 &   2.00 &    7.72 &   11.85 &    5.47 &    3.89 &    0.02 &    0.02 &    0.01 & 10.98 & 10.14\\
HAMS &   40 &   28 & M & 63.10 &   3.91 &    4.28 &    5.62 &    3.53 &    2.36 &    0.01 &    0.01 &    0.01 & 28.22 & 26.42\\
TAMS &    5 &  0.5 & S &  5.38 &   0.12 &   82.03 &  124027 &   83.84 &   22.09 &    0.52 &    0.66 &    0.79 &  0.92 &  0.93\\
TAMS &   10 &    1 & S & 10.83 &   0.15 &   19.82 &    8646 &   19.86 &    4.67 &    0.08 &    0.08 &    0.07 &  2.43 &  2.38\\
TAMS &   20 &    2 & S & 21.04 &   0.32 &    7.72 &     908 &    8.14 &    1.79 &    0.03 &    0.03 &    0.02 &  6.56 &  6.04\\
TAMS &   40 &    4 & S & 39.85 &   0.85 &    4.28 &     141 &    4.63 &    0.78 &    0.01 &    0.01 &    0.01 & 15.73 & 14.28\\
TAMS &    5 &    2 & S &  6.47 &   0.53 &   82.03 &     908 &   54.96 &   27.01 &    0.32 &    0.30 &    0.34 &  1.29 &  1.19\\
TAMS &   10 &    4 & S & 12.96 &   1.02 &   19.82 &     141 &   14.92 &    6.54 &    0.06 &    0.05 &    0.04 &  3.64 &  3.02\\
TAMS &   20 &    8 & S & 25.16 &   2.20 &    7.72 &   29.75 &    6.78 &    2.41 &    0.02 &    0.02 &    0.01 &  8.62 &  7.72\\
TAMS &   40 &   16 & S & 48.09 &   4.52 &    4.28 &    9.96 &    4.11 &    1.18 &    0.01 &    0.02 &    0.01 & 22.11 & 18.28\\
TAMS &    5 &  3.5 & P &  7.86 &   0.64 &   82.03 &     199 &   36.40 &    0.06 &    0.16 &   12.00 &   11.86 &  1.01 &  1.57\\
TAMS &   10 &    7 & P & 15.70 &   1.28 &   19.82 &   38.89 &   11.42 &    0.01 &    0.06 &    4.11 &    3.16 &  1.49 &  3.78\\
TAMS &   20 &   14 & M & 30.73 &   2.55 &    7.72 &   11.85 &    5.63 &    0.01 &    0.02 &    2.05 &    1.96 & 12.69 &  9.84\\
TAMS &   40 &   28 & M & 58.81 &   5.09 &    4.28 &    5.62 &    3.68 &    0.02 &    0.01 &    0.00 &    0.85 & 30.17 & 28.47\\
TAMS &    5 &    5 & M &  9.18 &   0.81 &   82.03 &   82.03 &   26.86 &    0.01 &    0.11 &    3.90 &    0.58 &  1.11 &  1.96\\
TAMS &   10 &   10 & M & 18.35 &   1.61 &   19.82 &   19.82 &    9.49 &    0.02 &    0.04 &    2.89 &    2.87 &  6.39 &  4.94\\
TAMS &   20 &   20 & M & 35.62 &   3.11 &    7.72 &    7.72 &    5.00 &    1.29 &    0.01 &    0.01 &    0.01 & 13.36 & 12.23\\
CHEX &    5 &  0.5 & S &  5.37 &   0.13 &   82.03 &  124027 &   84.23 &   23.28 &    0.61 &    0.63 &    0.77 &  0.92 &  0.93\\
CHEX &   10 &    1 & S & 10.81 &   0.17 &   19.82 &    8646 &   20.04 &    4.63 &    0.08 &    0.08 &    0.07 &  2.49 &  2.52\\
CHEX &   20 &    2 & S & 21.04 &   0.32 &    7.72 &     908 &    8.14 &    1.76 &    0.03 &    0.03 &    0.02 &  6.49 &  6.04\\
CHEX &   40 &    4 & S & 39.84 &   0.85 &    4.28 &     141 &    4.63 &    0.74 &    0.01 &    0.06 &    0.01 & 15.70 & 14.22\\
CHEX &    5 &    2 & M &  6.47 &   0.53 &   82.03 &     908 &   54.96 &   26.04 &    0.32 &    0.29 &    0.34 &  1.31 &  1.19\\
CHEX &   10 &    4 & S & 12.89 &   1.09 &   19.82 &     141 &   15.09 &    6.46 &    0.06 &    0.04 &    0.04 &  3.54 &  3.02\\
CHEX &   20 &    8 & S & 25.13 &   2.23 &    7.72 &   29.75 &    6.79 &    2.34 &    0.02 &    0.02 &    0.02 &  8.47 &  7.71\\
CHEX &   40 &   16 & S & 48.09 &   4.52 &    4.28 &    9.96 &    4.08 &    1.26 &    0.01 &    0.01 &    0.01 & 22.33 & 17.73\\
CHEX &    5 &  3.5 & P &  7.85 &   0.65 &   82.03 &     199 &    0.00 &    0.00 &    0.00 &   12.55 &   11.24 &  0.98 &  0.00\\
CHEX &   10 &    7 & P & 15.68 &   1.30 &   19.82 &   38.89 &   11.50 &    0.00 &    0.05 &    4.10 &    2.14 &  1.28 &  3.84\\
CHEX &   20 &   14 & P & 30.71 &   2.57 &    7.72 &   11.85 &    5.63 &    0.00 &    0.02 &    1.94 &    1.86 & 12.68 &  9.85\\
CHEX &   40 &   28 & M & 58.89 &   5.01 &    4.28 &    5.62 &    3.67 &    0.96 &    0.01 &    0.01 &    0.01 & 32.30 & 28.50\\
CHEX &    5 & 4.95 & M &  9.14 &   0.81 &   82.03 &     102 &   23.18 &    0.00 &    0.09 &    2.20 &    2.05 &  1.14 &  2.22\\
CHEX &    5 & 4.95 & M &  9.14 &   0.81 &   82.03 &     102 &   27.07 &    0.00 &    0.11 &    2.20 &    2.05 &  1.14 &  1.95\\
CHEX &   20 & 19.8 & M & 35.46 &   3.08 &    7.72 &    8.65 &    5.02 &    0.00 &    0.02 &    1.43 &    1.39 & 16.39 & 12.16\\
CHEX &   40 & 39.6 & M & 67.26 &   5.82 &    4.28 &    4.53 &    3.41 &    0.00 &    0.01 &    0.00 &    0.65 & 39.24 & 31.34\\

\end{tabular}
\end{minipage}
\end{table*}

For completeness Table \ref{tab:evolution_results} gives the total
remnant mass as well as the collision case. The table also lists the time
$t_\mathrm{ms}$ that the merger product spends as a core hydrogen
burning star and the time $t_\mathrm{HG}$ spend in the Hertzsprung gap,
between the end of core hydrogen burning (CHEX) and the base of the giant
branch. The duration of the hydrogen shell burning phase is given by
$t_\mathrm{HSB}$. This phase ends with the ignition of helium in the core.
Note that with these definitions the evolution in the Hertzsprung gap can
include part of the core-helium burning phase, in which case
$t_\mathrm{HG} > t_\mathrm{HSB}$. We discuss this in more detail in
section \ref{sec:evolution_no_hydrogen}. The table also gives the 
core mass $M_\mathrm{He}$ at helium ignition.

\subsubsection{Contraction phase and core hydrogen burning}
\label{sec:contraction_and_mainsequence_evolution}

\begin{figure}
\includegraphics[width=0.5\textwidth]{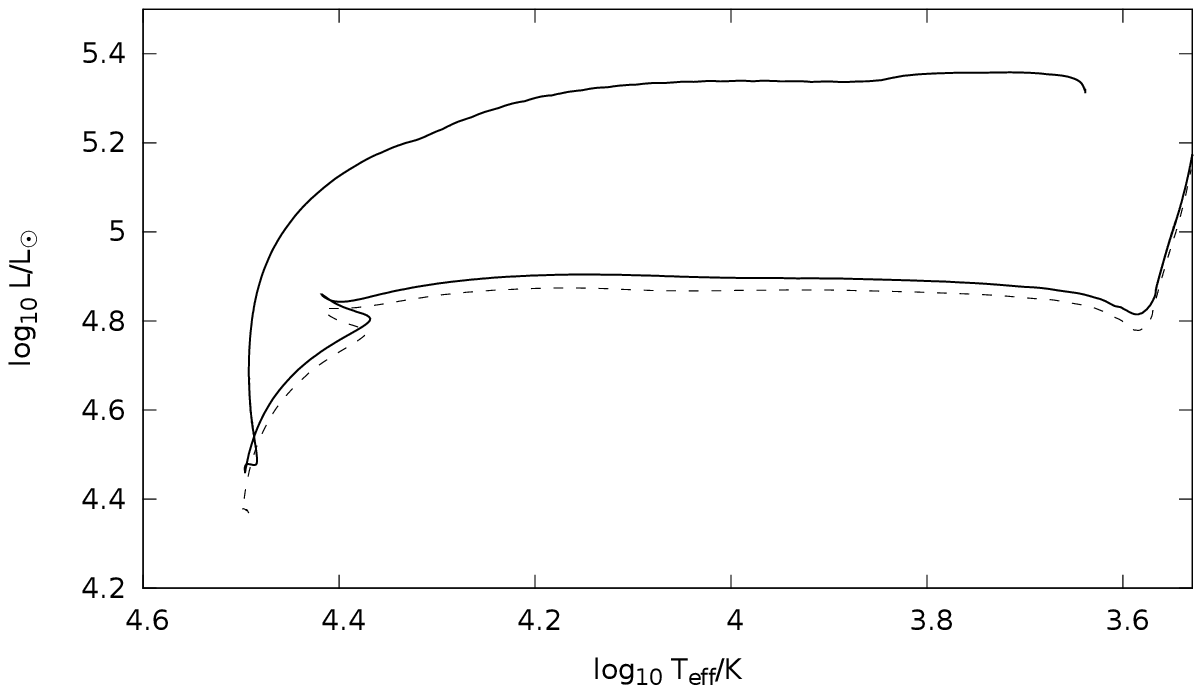}
\caption{Evolution track of the HAMS 10+7 merger product in the
Hertzsprung-Russel diagram (solid line). The dotted line shows the
evolution track of a normal star of the same mass as the merger
product.}
\label{fig:hrd_hams_10_7}
\end{figure}

\begin{figure}
\includegraphics[width=0.5\textwidth]{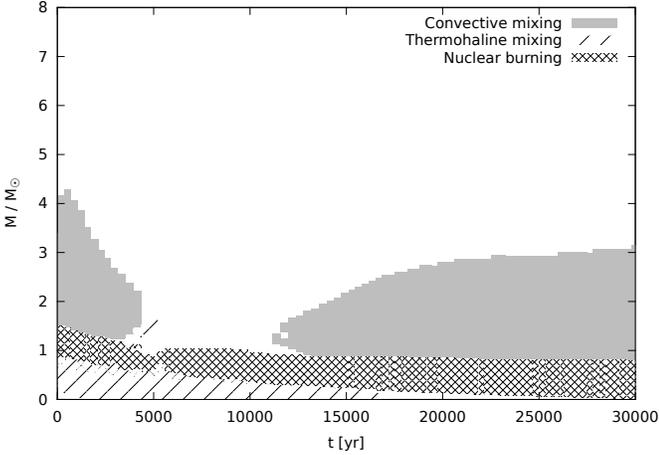}
\caption{Kippenhahn diagram showing the evolution of the HAMS 10+1
merger product during the contraction phase. Note that initially,
hydrogen is burning in a shell outside the core. Thermohaline mixing
is responsible for mixing into the central hydrogen rich core while
convection mixes the region just above the burning shell.}
\label{fig:kippenhahn_m10m1_hams}
\end{figure}

\begin{figure}
\includegraphics[width=0.5\textwidth]{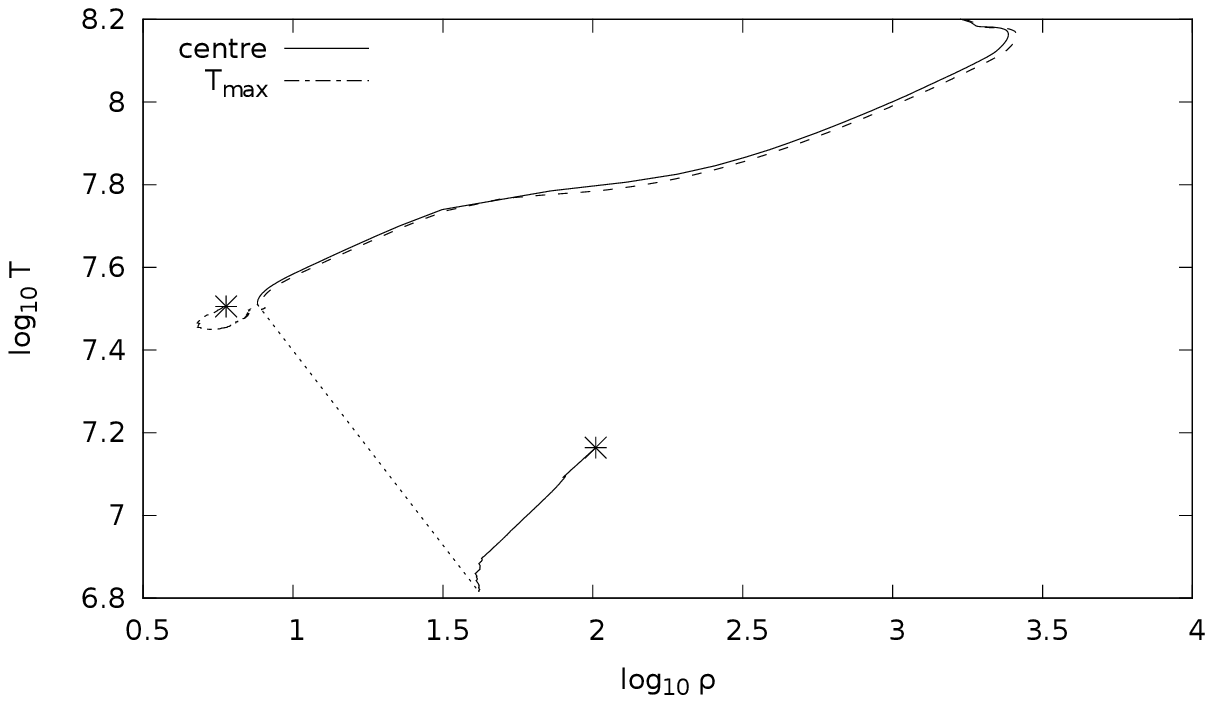}
\caption{Evolution track of the HAMS 10+1 merger product in the $\log
\rho$--$\log T$ plane. The solid line shows the evolution of the stellar
centre while the dash-dotted line shows the evolution of the locus of
highest temperature. The starting point of the evolution is indicated by
\plustimes. The dotted line shows the transition of hydrogen shell burning
to hydrogen core burning. For reference, the dashed line shows the
evolution track of a normal star of the same initial mass as the merger
product.}
\label{fig:rhot_hams_10_1}
\end{figure}

The first phase of evolution after the collision is the contraction phase,
which progresses similarly to that for low-mass merger products, as
studied in a previous paper \citep{GlebbeekPolsHurley2008} and before by other
authors \citep{sills_evcolprod1}. The evolution track for the HAMS 10+7
collision shown in Figure \ref{fig:hrd_hams_10_7} shows the typical
features for the merger product evolution tracks.
During the contraction phase the merger product is inflated and
over-luminous. In this example the stellar radius is inflated by a factor
of 100 compared to the equilibrium radius, but the star contracts quickly
and after only 200 years the radius is only five times the equilibrium radius.
The initial radius is sensitive to the assumptions made about the surface
layers (as discussed in Section \ref{sec:reduce_3d}).
Most of the star's luminosity is provided by the release of
thermal energy and the envelope contracts on a thermal timescale.
In all our merger products the envelope is enhanced in helium, which
means it is brighter than a normal star of the same mass, as we found for
low-mass mergers \citep{GlebbeekPolsHurley2008,2010MNRAS.408.1267G,GlebbeekPols2008}. 

If the core of the merger product is made up of material from the
primary (case P) or is a mixture of material from the progenitor stars
(case M), the highest temperature occurs in the centre of the merger
product. Because the core is hydrogen rich, hydrogen ignites in the centre.
Temperature inversions like the one found in the HAMS 10+1 collision
product can result in the formation of a hydrogen burning shell on top of a
hydrogen-rich core if the hydrogen abundance in the region of maximum
temperature is high enough.  During the contraction phase these temperature
inversions are removed and the burning front moves inward to the centre of
the remnant.  At the same time molecular weight inversions in the interior
are removed by thermohaline mixing \citep{kippenhahn_thermohalinemixing}.

The evolution of the stellar interior for the HAMS 10+1 merger product
(case S) is shown in Figure \ref{fig:kippenhahn_m10m1_hams} and Figure
\ref{fig:rhot_hams_10_1}. As discussed in section
\ref{sec:hams_structure} the core of the merger product consists of
material from the $1\,\msun$ secondary star and is overdense.  This means
that it needs to expand in order to reach thermal equilibrium. Because this
requires work against the pressure exerted by the envelope and the
expansion is nearly adiabatic the temperature in the core decreases. The
highest temperature occurs in a helium rich hydrogen burning shell above
the core.  Local conditions in the burning shell are close to the
prevailing conditions in the core of a normal $11\,\msun$ star (see the
dash-dotted line in Figure \ref{fig:rhot_hams_10_1}). The burning shell
becomes denser.  Because the molecular weight in the burning shell is
higher than the molecular weight in the core thermohaline mixing operates,
which slowly homogenises the inner part of the star.  Figure
\ref{fig:kippenhahn_m10m1_hams} shows the evolution in a Kippenhahn
diagram. The burning front moves gradually inward on a thermal timescale
while convection mixes the layers above the core.

The configuration of a contracting burning shell and an expanding core
(both on a thermal timescale) 
leads to an instability because it drives the core to a density inversion. 
We were able to follow the transition from hydrogen shell burning to
hydrogen core burning until the bottom of the burning shell was within
$0.03\,\msun$ of the centre. At this point we expect the transition from
shell burning to core burning to happen very shortly. The transition of
(stable) hydrogen shell burning to central hydrogen burning that we find
for some of our merger products is analogous to the transition of
(unstable) off-centre helium shell burning to helium core burning in the
helium flash.
Although we were not able to follow the transition to core burning
self-consistently, we can study the long term evolution of the merger
product by constructing a `post core ignition' model. This procedure is
similar to the procedure that is commonly used to model stellar
evolution after the helium flash (see
\citealt{2005A&A...442.1041S} for a description of this method
and comparison with more detailed calculations). In
Figure \ref{fig:rhot_hams_10_1} this transition is indicated by a dotted
line. After central hydrogen ignition the evolution of the merger
product is very similar to that of a normal star of the same mass
(dashed line), although the central density and pressure are a bit lower in the
merger product for the same central temperature.
The merger product has a slightly larger radius and the decrease in
central pressure is in agreement with the scaling relation $P_c \sim
GM^2/R^4$.

Two other interesting case S merger products are the CHEX 20+8 and 20+2
merger products (and the similar TAMS 20+8 and 20+2), described in
section \ref{sec:chex_structure}. In contrast to the HAMS 10+1 case, in
these cases no hydrogen burning shell is formed because the hydrogen
abundance at the location of maximum temperature is too low. Thermohaline
mixing very quickly homogenises the core. In neither of these cases do we
find significant mixing of the star due to convection outside the region of
the stellar core.  Although both merger remnants have a hydrogen rich core
of about $2\,\msun$, the inner $7\,\msun$ (corresponding to the extent of
the convective core) of the 20+2 remnant contains more helium than the
inner $10\,\msun$ of the 20+8 remnant. As a result and as can be seen from
Table \ref{tab:evolution_results} the 20+8 has a longer main sequence
lifetime than the 20+2. The same effect can be seen for 40+16 and 40+4.

Core hydrogen burning in these merger products proceeds normally and
these stars are very similar to normal stars of the same mass. After
core hydrogen exhaustion, hydrogen continues to burn in a shell. Because
the region of the burning shell can have an enhanced helium abundance
compared to a normal star of the same mass, the core mass at which helium
ignites in the centre is larger than in a normal star. As a result, the
shell burning phase (lifetime in the Hertzsprung gap, $t_\mathrm{HG}$ in
Table \ref{tab:evolution_results}) is slightly shorter than for a normal
star. 

\subsubsection{Merger products with hydrogen depleted cores}
\label{sec:evolution_no_hydrogen}

\begin{figure}
\includegraphics[width=0.5\textwidth]{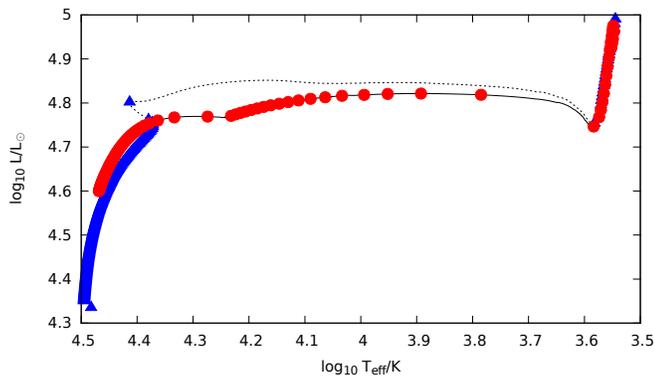}
\caption{Evolution track of the TAMS 10+7 merger product (solid line)
compared to the evolution track of a normal star of the same mass (dotted
line). Points are plotted along the curves every $50\,000$ years. Note that
the merger product spends a considerably longer time in the blue part of
the region corresponding to the Hertzsprung gap than the normal star of the
same mass.}
\label{fig:hrd_tams_10_7}
\end{figure}

\begin{figure}
\includegraphics[width=0.5\textwidth]{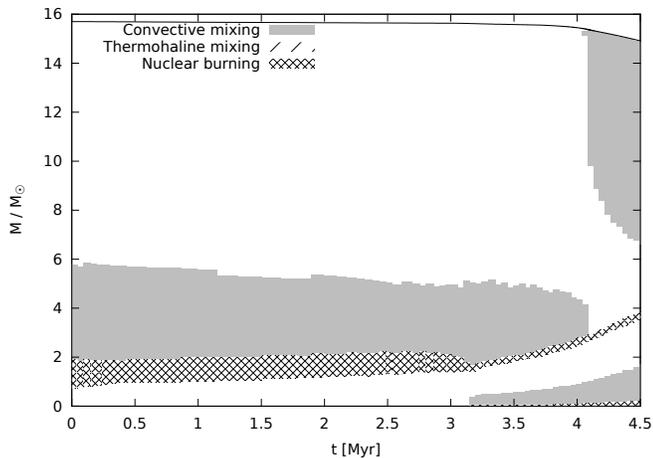}
\caption{Kippenhahn diagram showing the evolution of the 10+7 \msun TAMS
merger product. Hydrogen burns in a shell around the helium core.
}
\label{fig:kippenhahn_m10m7_tams}
\end{figure}

\begin{figure}
\includegraphics[width=0.5\textwidth]{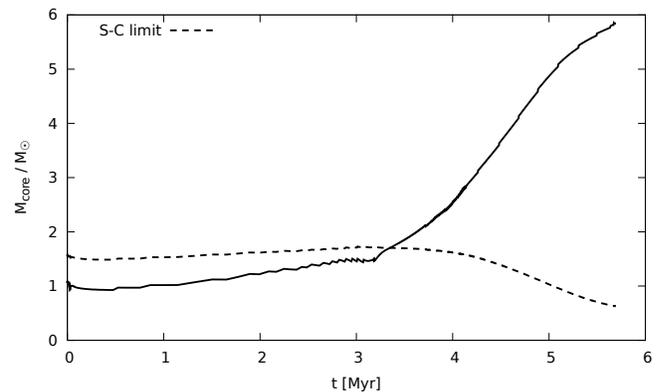}
\caption{Core mass as functions of time for the TAMS 10+7 merger,
in solar units.  The dashed line represents the Sch\"onberg-Chandrasekhar
limit for the core mass. The merger product ignites helium after 3.1
Myr.
} \label{fig:sc_limit}
\end{figure}

\begin{figure}
\includegraphics[width=0.5\textwidth]{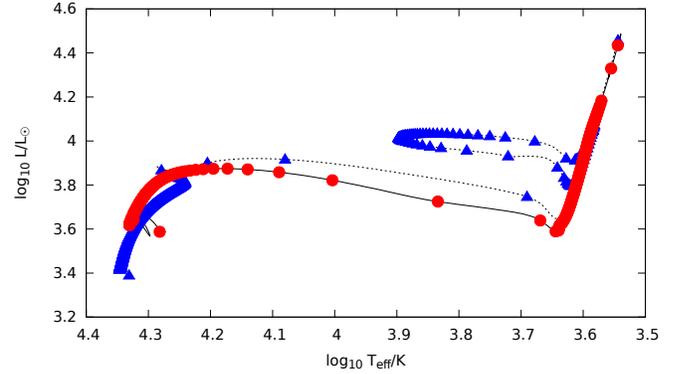}
\caption{As Figure \ref{fig:hrd_tams_10_7} for the TAMS 5+3.5 merger.}
\label{fig:hrd_tams_5_35}
\end{figure}

If the merger product has a hydrogen-depleted core (case P for TAMS or
CHEX merger and some case M for TAMS or CHEX merger with $q \ge
0.99$) the evolution track can be very different from that of a normal star
of the same mass or the merger products discussed in
\ref{sec:contraction_and_mainsequence_evolution}. These merger products
do not have a core hydrogen-burning phase and begin their evolution 
with hydrogen shell burning on top of a helium core. After hydrogen
exhaustion a normal star will quickly evolve through the Hertzsprung gap to
the giant branch. Some of the merger products, however, have a
long-lived phase of hydrogen shell burning in the blue part of the CMD and
may even spend a significant part of their core helium burning lifetime
while blue (Figures \ref{fig:hrd_tams_10_7} and \ref{fig:hrd_tams_5_35}).

Figure \ref{fig:hrd_tams_10_7} shows the evolution track of the TAMS 10+7
merger product. The location in the Hertzsprung-Russell diagram after
the contraction phase is similar to that of a normal star of the same mass
late on the main sequence but the evolution track lacks the
distinctive hook that appears when the convective core disappears in a
normal star. Helium ignition occurs in the Hertzsprung gap near $\log_{10}
T_\mathrm{eff} = 4.3$.
The evolution of the interior is shown in Figure \ref{fig:kippenhahn_m10m7_tams}.
The hydrogen burning shell is replenished by the thick convection zone
above it.  The extent of this convection zone is comparable to the extent
of the convective core in a normal star of the same mass. As a result, the
hydrogen burning shell mimics the hydrogen core of a normal star. 
Because of the presence of a convection zone above the burning shell the
core mass grows slowly, increasing by only $0.5\,\msun$ in the first 3.1
Myr. After 3.1 Myr helium ignites in the centre. The convection zone that
sustained the burning shell shrinks but does not disappear. This causes
the core growth rate to increase after helium ignition.

An interesting aspect of this is that although the merger product is
not a main sequence star, it would appear on the extension of the
main sequence in a star cluster, and be counted as a blue straggler star.
This is an exception to the result of \citet{1997ApJ...484L..51S} that mergers
involving stars with hydrogen depleted cores do not produce blue
stragglers.

The question of what allows a star to have an extended blue phase of
hydrogen shell burning and what causes it to become a red giant instead has
been debated extensively in the literature \citep{1981ASSL...88..179E,
1985ApJ...296..554Y, 1988ApJ...329..803A, 1989A&A...209..135W,
1991ApJ...383..757E, 1992ApJ...400..280R, 2000ApJ...538..837S,
2009PASA...26..203S, 2012MNRAS.421.2713B}
and it is not clear that there is a single answer.

For hydrogen shell burning stars with an isothermal core the
transition can be understood in terms of the maximum core mass, set by the
virial theorem, for which the core can avoid contraction. While the core
mass is below this limit the star remains in the blue part of the
Hertzsprung-Russell diagram.
This is the Sch\"onberg-Chandrasekhar (SC) limit
\citep{1942ApJ....96..161S,kippenhahn_weigert,2012MNRAS.421.2713B},
\begin{equation}\label{eqn:sc}
M_\mathrm{SC} = 0.37 \left(\frac{\mu_\mathrm{env}}{\mu_\mathrm{c}}\right)^2
M.
\end{equation}
Here $M$ is the total mass of the star, $\mu_\mathrm{env}$ the average mean
molecular weight of the envelope and $\mu_\mathrm{c}$ the average mean
molecular weight in the core. As long as the core mass is below the SC
limit, the star can have an isothermal core and remain in thermal
equilibrium.
For normal stars in the mass range we consider here, the convective core on
the main sequence is already more massive than this limit. However, this is
not the case for the merger products.

In Figure \ref{fig:sc_limit} we show the core mass of the TAMS 10+7 merger
product as a function of time.
The dashed line shows the SC limit as calculated by Eq. (\ref{eqn:sc}). The
core mass is below this limit until $3.1\,\mathrm{Myr}$.
As the core mass approaches the SC limit the core contracts rapidly and
heats up, leading to helium ignition. 
At this point the merger product is still in the blue part of the
Hertzsprung-Russell diagram and becomes a blue supergiant, as can be seen
from Figure \ref{fig:hrd_tams_10_7}.
In Table \ref{tab:evolution_results} the merger products that form blue
supergiants all have a helium core mass $M_\mathrm{He}$ smaller
than the core mass $M_\mathrm{He, ms}$ of a normal star and
$t_\mathrm{HSB}$ much smaller than $\tau_\mathrm{HG}$.

In addition to a prolonged blue phase of hydrogen shell burning
some of our merger models also have a blue phase of helium core burning.
The classical SC limit only applies to isothermal
cores and is no longer relevant once a temperature gradient develops in the
core and it cannot explain why some of these stars also stay blue during
core helium burning. It is tempting to
relate the occurrence of a blue phase of core helium burning to the
occurrence of blue loops, which with our evolution code occur for stars
below 12 $M_\odot$.
The occurrence of blue loops has been linked to
the efficiency of the hydrogen burning shell \citep{XuLi2004}
but as has been noted already by \citep{1971A&A....10...97L}, the
presence of blue loops depends sensitively on the hydrogen profile in the
envelope. Since the abundenace profile of our merger models can be very
different from that of normal stars it is not surprising that the blue
loops can be very different.

Not all stars with $M_\mathrm{He} < M_\mathrm{He, ms}$ form blue
supergiants, however. The remnant of the TAMS 5+3.5 merger, shown in
Figure \ref{fig:hrd_tams_5_35}, also stays in the blue part of the
Hertzsprung-Russell diagram during a prolonged hydrogen shell burning
phase, but in this case there is no blue phase of core helium burning. This
is actually different from the main sequence reference model.
In the reference model, helium ignition occurs near the tip of the giant
branch ($L \approx 4.0 L_\odot$) and the star then makes a blue loop before
evolving up the AGB. The merger ignites helium in the Hertzsprung gap,
before reaching the base of the giant branch, and continues to evolve up
the giant branch until he exhaustion at $L \approx 4.2 L_\odot$. It then
evolves up the AGB.

\subsubsection{An analytical recipe}
\begin{figure}
\includegraphics[width=0.5\textwidth]{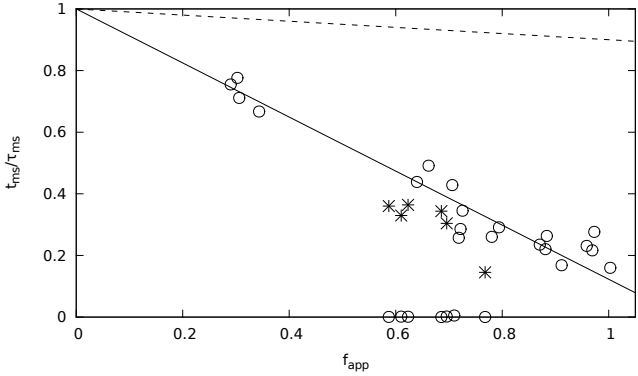}
\caption{Collision product lifetime compared to the prediction from
Eq. (\ref{eqn:apparent_lifetime}). The models are indicated by $\circ$, the
solid line is the prediction of Eq. (\ref{eqn:apparent_lifetime}) for
$\alpha=1.14$. The dotted line is the
lifetime according to the prescription of
\citet{hurley_sse}. Models with no main sequence but with an extended blue
phase of hydrogen shell burning are plotted with $\plustimes$ for the
duration of this phase.}
\label{fig:lifetime}
\end{figure}

An analytic recipe that can be used to predict merger product lifetimes
and luminosities was developed by \citet{GlebbeekPols2008}
In brief, they found that the lifetime $t_\mathrm{MS}$ of the
merger product can be found from $t_\mathrm{MS} = \tau_\mathrm{MS}
(1-f_\mathrm{app}/\alpha)$, where 
\begin{equation}\label{eqn:apparent_lifetime}
f_\mathrm{app} = 
\frac{1}{Q_\mathrm{c}(M)}
\frac{1}{1-\phi}
\frac{ Q_{\mathrm{c},1} f_1 + Q_{\mathrm{c},2} f_2 q} { 1+q }.
\end{equation}
Here $\tau_\mathrm{MS}$ is the main sequence lifetime of a star of the same
mass as the merger product, $f_1$ and $f_2$ are the ages of the primary
and secondary star at the time of the merger, in units of their main sequence lifetimes, $\phi$ is
the fraction of material lost in the collision, $Q_\mathrm{c}(M)$ is the
fraction of hydrogen that is consumed during the main sequence by a
star of mass $M$ \citep{GlebbeekPols2008}, $M$ is the total mass
of the merger, $Q_{\mathrm{c},1}$ and $Q_{\mathrm{c},2}$ are the
fractions of hydrogen consumed in the primary and secondary during their
lifetime and $\alpha$
is a free parameter. \citet{GlebbeekPols2008} found that a single value of
$\alpha=1.67$ works well for mergers of low-mass stars. We find that
$\alpha = 1.14$ works well for the high-mass mergers discussed
here (see Figure \ref{fig:lifetime}).

Some of our models have no main-sequence phase because they begin with
hydrogen depleted cores. In Figure \ref{fig:lifetime} these models
fall along the bottom of the plot. Some of these have an extended blue phase of
hydrogen shell burning, and we also plot these with the duration of the
hydrogen shell burning phase in place of $t_\mathrm{ms}$. They then fall
close to the predicted main-sequence lifetime.

\subsection{Surface composition}

\begin{figure}
\includegraphics[width=0.5\textwidth]{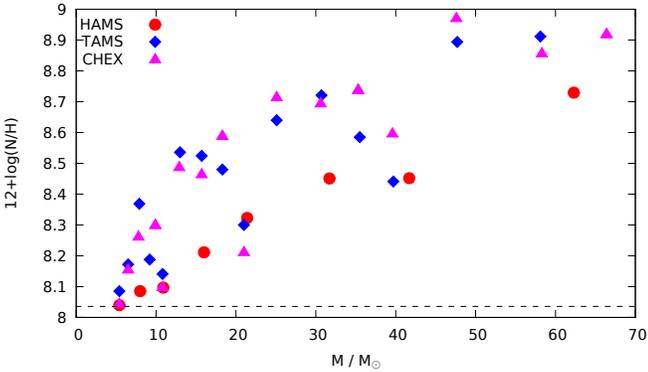}
\caption{Surface abundance of nitrogen for our merger models, as a function
of total mass. The dashed line represents our ZAMS composition.}
\label{fig:surface_comp}
\end{figure}

\begin{table}
\caption{Surface abundances (by mass fraction) for the different merger
products. The ZAMS line indicates the assumed ZAMS composition. The given
abundance of CNO is multiplied by 1000.}
\label{tab:surface_comp}
\begin{tabular}{lrrrrrrr}
\hline
stage & $M_1$ & $M_2$ & $X_\mathrm{H}$ & $X_\mathrm{He}$ & $X_\mathrm{C}$ & $X_\mathrm{N}$ & $X_\mathrm{O}$\\
& \msun & \msun & & & $\times 10^{3}$& $\times 10^{3}$&
$\times 10^{3}$\\
\hline
ZAMS &   &     & 0.700 & 0.280 &  3.52 & 1.04 & 10.04\\
HAMS & 5 & 0.5 & 0.700 & 0.280 &  3.50 & 1.07 & 10.03\\
HAMS & 10 & 1 & 0.699 & 0.281 &  3.43 & 1.22 & 9.95\\
HAMS & 20 & 2 & 0.690 & 0.290 &  3.18 & 2.03 & 9.36\\
HAMS & 40 & 4 & 0.683 & 0.297 &  3.04 & 2.71 & 8.77\\
HAMS & 5 & 3.5 & 0.700 & 0.280 &  3.43 & 1.19 & 9.99\\
HAMS & 10 & 7 & 0.698 & 0.282 &  3.27 & 1.59 & 9.74\\
HAMS & 20 & 14 & 0.691 & 0.289 &  2.87 & 2.73 & 8.97\\
HAMS & 40 & 28 & 0.662 & 0.318 &  2.39 & 4.97 & 7.06\\
TAMS & 5 & 0.5 & 0.700 & 0.280 &  3.42 & 1.19 & 10.00\\
TAMS & 10 & 1 & 0.697 & 0.283 &  3.37 & 1.35 & 9.88\\
TAMS & 20 & 2 & 0.695 & 0.285 &  3.13 & 1.94 & 9.53\\
TAMS & 40 & 4 & 0.682 & 0.298 &  3.05 & 2.64 & 8.85\\
TAMS & 5 & 2 & 0.698 & 0.282 &  3.32 & 1.45 & 9.87\\
TAMS & 10 & 4 & 0.689 & 0.291 &  2.60 & 3.31 & 8.70\\
TAMS & 20 & 8 & 0.668 & 0.312 &  2.44 & 4.08 & 8.01\\
TAMS & 40 & 16 & 0.625 & 0.355 &  1.85 & 6.85 & 5.63\\
TAMS & 5 & 3.5 & 0.694 & 0.286 &  2.83 & 2.27 & 9.55\\
TAMS & 10 & 7 & 0.691 & 0.289 &  2.53 & 3.23 & 8.86\\
TAMS & 20 & 14 & 0.658 & 0.322 &  2.16 & 4.85 & 7.51\\
TAMS & 40 & 28 & 0.615 & 0.365 &  1.81 & 7.03 & 5.50\\
TAMS & 5 & 5 & 0.697 & 0.283 &  4.73 & 1.50 & 8.39\\
TAMS & 10 & 10 & 0.687 & 0.293 &  3.11 & 2.90 & 8.59\\
TAMS & 20 & 20 & 0.679 & 0.301 &  2.47 & 3.66 & 8.45\\
CHEX & 5 & 0.5 & 0.700 & 0.280 &  3.66 & 1.09 & 9.86\\
CHEX & 10 & 1 & 0.699 & 0.281 &  3.42 & 1.23 & 9.95\\
CHEX & 20 & 2 & 0.696 & 0.284 &  3.31 & 1.59 & 9.69\\
CHEX & 40 & 4 & 0.672 & 0.308 &  2.74 & 3.72 & 8.02\\
CHEX & 5 & 2 & 0.697 & 0.283 &  3.32 & 1.40 & 9.90\\
CHEX & 10 & 4 & 0.690 & 0.290 &  2.65 & 2.97 & 8.99\\
CHEX & 20 & 8 & 0.665 & 0.315 &  2.16 & 4.83 & 7.53\\
CHEX & 40 & 16 & 0.619 & 0.362 &  1.50 & 8.12 & 4.66\\
CHEX & 5 & 3.5 & 0.696 & 0.284 &  3.13 & 1.78 & 9.72\\
CHEX & 10 & 7 & 0.691 & 0.289 &  2.70 & 2.82 & 9.10\\
CHEX & 20 & 14 & 0.665 & 0.315 &  2.24 & 4.61 & 7.67\\
CHEX & 40 & 28 & 0.634 & 0.346 &  1.97 & 6.40 & 6.00\\
CHEX & 5 & 5 & 0.694 & 0.286 &  4.51 & 1.94 & 8.19\\
CHEX & 10 & 9.9 & 0.677 & 0.303 &  2.43 & 3.68 & 8.48\\
CHEX & 20 & 19.8 & 0.649 & 0.331 &  2.17 & 4.98 & 7.37\\
CHEX & 40 & 39.6 & 0.589 & 0.391 &  2.22 & 6.86 & 5.25\\
CHEX & 5 & 5 & 0.694 & 0.286 &  4.51 & 1.94 & 8.19\\
CHEX & 10 & 9.9 & 0.677 & 0.303 &  2.43 & 3.68 & 8.48\\
CHEX & 20 & 19.8 & 0.649 & 0.331 &  2.17 & 4.98 & 7.37\\
CHEX & 40 & 39.6 & 0.589 & 0.391 &  2.22 & 6.86 & 5.25\\
\hline
\end{tabular}
\end{table}

Although there is generally little or no mixing of hydrogen into the core
of the merger product, the surface abundances can be strongly affected by
material from the secondary star that is mixed into the envelope. In
Table \ref{tab:surface_comp} we give the surface abundance (by mass
fraction) for the main elements (H, He, C, N and O) and in Figure
\ref{fig:surface_comp} we plot the surface nitrogen abundance as a function
of total remnant mass. In both cases we list the abundances once the merger
product has reached equilibrium and thermohaline convection has been
allowed to modify the surface abundances.

Our lowest mass models show the least amount of mixing, in agreement with
earlier findings in studies of mergers between low mass stars
\citep{1996ApJ...468..797L,sills_evcolprod1,GlebbeekPolsHurley2008}. In
contrast, our most massive merger products show much stronger mixing.
The strongest enhancement is found for merger products resulting from more
evolved parent stars.

It is clear from these results that substantial mixing can result from the
merger process itself. In fact, the range of nitrogen enhancement is
comparable for the range reported for B stars in the galactic field
\citep{1992A&A...262..171K,2012A&A...539A.143N} and in the Magellanic
Clouds \citep{2007A&A...466..277H}. It should be remembered, however, that
the initial initial composition of our models is different from the derived
compositions in these works and so our results are not directly comparable.

\section{Summary \& Conclusion}\label{sect:discussion}
We have studied the evolution of stellar mergers formed by a collision
involving massive stars. This is a first step in following the evolution of
stellar merger runaways using detailed stellar models.
Mass loss from the collisions is generally small, up to about 8\% of the
total mass for collisions involving equal mass stars at the end of the main
sequence.
The structure of these merger remnants can be well understood using a
modification of the entropy sorting principle of \citet{2002ApJ...568..939L}
presented by \citet{2008MNRAS.383L...5G}.

During the collision, the core of one of the parent stars can retain its
identity and sink to the centre of the merger product. Whether this
occurs for the core of the primary or the secondary depends on the buoyancy
profile $A(m)$ in the parent stars. In cases with extreme mass ratio the
secondary star as a whole can migrate to the centre of the merger
product.
In general we find that little hydrogen is mixed into the core of
the merger product. However, the merger product can still have a hydrogen
rich core if the core of the secondary displaces the core of the primary at
the centre of the merger product.
The shell with the location of highest temperature can initially be
off-centre, resulting in a hydrogen-burning shell in the merger remnant.
The burning front moves inward towards the centre on a thermal timescale.
During the merger processed material can be mixed into the envelope
of the merger product and produce strong nitrogen enhancement at the
surface, especially for mergers involving evolved parent stars.

The main-sequence evolution of the merger remnant is qualitatively similar
to the evolution of low mass collision products
\citep{sills_evcolprod1,sills_evcolprod2,GlebbeekPolsHurley2008,GlebbeekPols2008}:
in both cases helium enhancement of the lower envelope increases the star's
radius and luminosity. This increases the cross section of the merger
remnant for a possible subsequent collision, which may be relevant when
considering merger runaways. The main-sequence lifetime of massive
merger products can be predicted in a similar manner to that of low mass
merger products \citep{GlebbeekPols2008}.

Merger remnants that result from collisions with main-sequence stars at the
end of the main sequence and where the core of the primary remains intact
(case P) begin their evolution with an anomalously small
hydrogen-depleted core and evolve differently from normal main-sequence stars
or merger remnants that form with a hydrogen-rich core. The merger
products can be in thermal equilibrium during hydrogen shell burning if
the mass of the helium core is below the Sch\"onberg-Chandrasekhar limit.
In this case the star can appear as a blue straggler and ignite helium
while still in the blue part of
the Hertzsprung-Russell diagram. Observationally, many O and B stars appear
to be in the Hertzsprung gap \citep[][]{2006A&A...456..623E}, between
the end of the main sequence and the bluest part of the blue loop
during core helium burning, although this conclusion is sensitive to the
treatment of convective overshooting in the evolution tracks used for
comparison.
Stellar mergers involving turn-off stars, either through a collision as
discussed in this work or by unstable case-B mass transfer in a binary,
offer a mechanism to explain the presence of at least some stars in this
part of the Hertzsprung-Russell diagram.

In this work we have manually coupled the stellar evolution
calculations and the SPH calculations and performed a series of
collision and subsequent evolution experiments.  We coupled these
codes by hand, because there was not yet an automated way to do this
by the time we performed this research. The methods of converting
stellar evolution code output to SPH realizations and vice versa, as
described in this paper, have now been incorporated in the
Astronomical Multipurpose Software Environment (AMUSE for short, see
\citealt{2009NewA...14..369P,2013CoPhC.183..456P}).  AMUSE is a general-purpose
framework for interconnecting scientific simulation programs using a
homogeneous, unified software interface, and incorporates codes for
stellar evolution, hydrodynamics, gravity and radiative transport.
The entire software package including the scripts to perform the
calculations in this paper can be downloaded for free from {\tt
http://amusecode.org}.

\begin{acknowledgements}
We thank the referee, Georges Meynet, for useful comments and suggestions.
EG is supported by NWO under VENI grant 639.041.129.
\end{acknowledgements}

\bibliographystyle{mnras}
\bibliography{high_mass_grid}

 \end{document}